\documentclass{article}
\usepackage{amsmath,amsfonts,amsthm,amssymb,amscd,color,xcolor,mathrsfs}
\usepackage{hyperref}

\colorlet{darkblue}{blue!50!black}

\hypersetup{
    colorlinks,%
    citecolor=darkblue,%
    filecolor=red,%
    linkcolor=darkblue,%
    urlcolor=magenta,%
    pdfnewwindow=true,%
    pdfstartview={FitH}
}

\newcommand{\p}{\partial}
\newcommand{\e}{\varepsilon}

\newcommand{\R}{{\mathbb R}}
\newcommand{\Z}{{\mathbb Z}}
\newcommand{\IP}{{\mathbb P}}

\newcommand{\E}{{\mathbb E}}
\newcommand{\Emn}{{\mathbb E}^{\mu^\nu}}
\newcommand{\T}{{\mathbb T}}

\newcommand{\N}{{\mathbb N}}

\newcommand{\aA}{{\mathcal A}}
\newcommand{\BB}{{\mathcal B}}

\newcommand{\DD}{{\mathcal D}}
\newcommand{\EE}{{\mathcal E}}
\newcommand{\FF}{{\mathcal F}}

\newcommand{\HH}{{\mathcal H}}
\newcommand{\II}{{\mathcal I}}

\newcommand{\NN}{{\mathcal N}}
\newcommand{\OO}{{\mathcal O}}
\newcommand{\PP}{{\mathcal P}}

\newcommand{\VV}{{\mathcal V}}

\newcommand{\XX}{{\mathcal X}}

\newcommand{\dd}{{\textup d}}

\newcommand{\PPPP}{{\mathfrak P}}

\newcommand{\BBBB}{{\mathfrak B}}

\newcommand{\rot}{\mathop{\rm rot}\nolimits}
\newcommand{\Vol}{\mathop{\rm Vol}\nolimits}
\newcommand{\const}{\mathop{\rm const}\nolimits}

\newcommand{\supp}{\mathop{\rm supp}\nolimits}
\newcommand{\diver}{\mathop{\rm div}\nolimits}

\theoremstyle{plain}

\newtheorem{theorem}{Theorem}[section]

\theoremstyle{definition}
\newtheorem{definition}[theorem]{Definition}

\newtheorem{OP}{Open problem}
\theoremstyle{remark}

\newtheorem{example}[theorem]{Example}

\numberwithin{equation}{section}

\begin{document}

\author{Sergei Kuksin$^{\text{a}}$ and Armen Shirikyan$^{\text{b,c}}$}
\date{\small $^{\text{a}}$\,CNRS, Institut de Math\'emathiques de Jussieu--Paris Rive Gauche, UMR 7586, Universit\'e Paris Diderot, Sorbonne Paris Cit\'e, F-75013, Paris, France\\ 
E-mail: \href{mailto:Sergei.Kuksin@imj-prg.fr}{Sergei.Kuksin@imj-prg.fr}\\[5pt]
$^{\text{b}}$\,D\'epartement de Math\'ematiques, Universit\'e de Cergy--Pontoise, 
CNRS UMR8088\\
2 avenue Adolphe Chauvin, 95302 Cergy--Pontoise Cedex, France\\ 
E-mail: \href{mailto:Armen.Shirikyan@u-cergy.fr}{Armen.Shirikyan@u-cergy.fr}\\[5pt]
$^{\text{c}}$\,Centre de Recherches Math\'ematiques, CNRS UMI3457, Universit\'e de Montr\'eal\\ 
Montr\'eal,  QC, H3C 3J7, Canada\/}

\title{Rigorous results in space-periodic two-dimensional turbulence}
\maketitle

\begin{abstract}
We survey the  recent advance in the rigorous qualitative theory of the 2d stochastic Navier--Stokes system that are relevant to the description of turbulence in two-dimensional fluids. After discussing briefly the initial-boundary value problem and the associated Markov process, we formulate  results on the existence, uniqueness and mixing of a stationary measure. We next turn to various consequences of these properties: strong law of large numbers, central limit theorem, and random attractors related to a unique stationary measure. We also discuss the Donsker--Varadhan and Freidlin--Wentzell type large deviations, as well as the inviscid limit and asymptotic results in 3d thin domains. We conclude with some open problems.
\smallskip


\end{abstract}

\tableofcontents

\setcounter{section}{-1}

\section{Introduction}
\label{s0} 

The main subject of this article is the two-dimensional Navier--Stokes system in~$\R^2$ subject to periodic boundary conditions, perturbed by a random force. 
We thus consider the equations
\begin{equation} \label{0.1}
\p_tu+\langle u,\nabla \rangle u-\nu\Delta u+\nabla p=f(t,x), \quad\diver u=0,
\end{equation}
where $u=(u_1,u_2)$  and~$p$ are the velocity field and pressure of a fluid, $\nu>0$ is the kinematic viscosity, $f$ is an external (random) force, and $\langle u,\nabla \rangle=u_1\p_1+u_1\p_2$. All the functions are assumed to be $2\pi$-periodic in the spacial variables~$x_1$ and~$x_2$. Equations~\eqref{0.1} are supplemented with the initial condition
\begin{equation} \label{0.2}
u(0,x)=u_0(x),
\end{equation}
where $u_0$ is a given divergence-free vector field that is $2\pi$-periodic and locally square-integrable. Our aim is to review 
rigorous results on the qualitative behaviour of solutions for \eqref{0.1}
 as $t\to\infty$ and/or $\nu\to0$. These two limits are important for the mathematical description of the space--periodic 
2d turbulence (for  a   physical treatment of  this topic see e.g.~\cite{BV-2012}).  In our review we avoid detailed discussion, related to the relevance of the 2d  Navier--Stokes system~\eqref{0.1}  for    physics,      referring the reader to~\cite{batchelor1982,frisch1995,gallavotti2002, BV-2012}. But we   mention that  a number of 
  equations, closely related to the 2d Navier--Stokes system, are used in meteorology and oceanography, 
   so  the methods, developed for Navier--Stokes equations, can be applied in the mathematical 
   theory of climate and the statistical    description of the ocean.  A result, presented in Section~\ref{s6}, gives
    a very basic explanation for the relevance of the 2d models for statistical description of the 
     3d phenomena. 
  
Most of the results, discussed in our work, were obtained in this century. Complete proofs and  discussion  of many of them can be found  in~\cite{KS-book}. New material, not treated in that book, includes  the theory of large deviations and recent progress concerning the mixing in Eq.~\eqref{0.1}. 
\smallskip

\noindent 
{\bf Acknowledgement.}
This research was supported by the {\it Agence Nationale de la Recherche\/} through the grants ANR-10-BLAN~0102 and ANR-17-CE40-0006-02. The research of AS was carried out within the MME-DII Center of Excellence (ANR-11-LABX-0023-01) and supported by the {\it Initiative d'excellence Paris-Seine\/} and by the CNRS PICS {\it Fluctuation theorems in stochastic systems\/}.

\section{Equations and random forces}
\label{s1} 
\subsection{Cauchy problem}
\label{s1.1}
As was mentioned in the Introduction, we consider Eqs.~\eqref{0.1} with periodic boundary conditions. Before describing the class of random forces~$f$ we deal with, let us recall a general result on the existence, uniqueness, and regularity of solutions to the deterministic 
Cauchy problem~\eqref{0.1},~\eqref{0.2} We begin with the definition of a solution on an arbitrary time interval~$J_T=[0,T]$. The definitions of all the functional spaces used here and henceforth can be found in the list of frequently used notations at the end of the paper. 

\begin{definition} \label{d1.1}
Let $f(t,x)$ be the time derivative of a piecewise continuous function\footnote{Here and in similar situations below, it means that $g$ has at most countably many points of discontinuity, where it  has left and right limits. Traditionally we normalise such functions to be right-continuous everywhere. In particular, $g(0)$ is well defined.} $g:J_T\to L^2(\T^2,\R^2)$, vanishing at $t=0$: $f=\p_t g(t,x)$. A function $u(t,x)$ defined on~$J_T\times\T^2$ is called a {\it weak solution\/} for~\eqref{0.1} if it belongs to the space 
$$
\XX_T=C(J_T,L_\sigma^2)\cap L^2(J_T,H^1(\T^2,\R^2)\cap  L^2 _\sigma),
$$
and satisfies the relation
\begin{equation} \label{1.1}
\bigl(u(t),\varphi\bigr)
+\int_0^t\bigl(\langle u,\nabla \rangle u-\nu\Delta u,\varphi\bigr)\dd s
=\bigl(u(0),\varphi\bigr)+\bigl(g(t),\varphi\bigr), \quad t\in J_T,
\end{equation}
where $\varphi$ is an arbitrary divergence-free  smooth vector field on~$\T^2$, and the term with the Laplacian under the integral is understood in the weak sense: $(\Delta u,\varphi)=-(\nabla u,\nabla\varphi)$. In what follows, when discussing Eqs.~\eqref{0.1}, we often say {\it solution\/} rather than {\it weak solution}.
\end{definition}

Note that the Navier--Stokes system contains three unknown functions, $u_1$, $u_2$, and~$p$, whereas the definition of a solution specifies only the velocity field~$u$. This is due to the fact that, once $u(t,x)$ satisfying~\eqref{1.1} is constructed, the Leray decomposition (see Theorem~1.5 in~\cite[Chapter~1]{temam1979}) may be used to find a  unique (up to an additive function of~$t$) distribution $p(t,x)$ such that the first equation in~\eqref{0.1} holds in a weak sense. A proof of the following theorem is essentially contained in~\cite[Chapter~3]{temam1979}.

\begin{theorem} \label{t1.2}
Let $T>0$ and let $f$ be the sum of a square-integrable function $h(x)$ and the time derivative of a piecewise continuous function with range in~$H^1(\T^2,\R^2)$. Then, for any $u_0\in L_\sigma^2$, Eq.~\eqref{0.1} has a unique weak solution $u\in\XX_T$ satisfying the initial condition~\eqref{0.2}.  
\end{theorem}

If the function $g(t,x)$ is such that its mean value (in~$x$) vanishes identically in $t$, then the mean value of the solution $u(t,x)$ is time-independent. Below we always assume that the mean values of the forces we apply to the Navier--Stokes system and of its solutions which we consider, both vanish identically in time. 

Theorem~\ref{t1.2} allows one to construct a unique solution of Eq.~\eqref{0.1}  for the three important classes of random forces $f$, specified below. Namely, let~$H$ be the space of divergence free  square-integrable vector fields on~$\T^2$ with zero mean value. We shall deal with random forces of the form
\begin{equation} \label{1.2}
f(t,x)=h(x)+\eta(t,x),
\end{equation}
where $h\in H$ is a deterministic function, and~$\eta$ is one of the following three random processes:

\medskip
\noindent
{\bf Spatially regular white noise.}
Let $\Z_*^2$ be the set of non-zero integer vectors $j=(j_1,j_2)$ and let $\{e_j,j\in\Z_*^2\}$ be a trigonometric basis in~$H$ defined by 
\begin{equation} \label{1.3}
e_j(x)=\frac{j^\bot}{\sqrt{2}\,\pi |j|}\left\{
\begin{aligned}
\cos\langle j,x\rangle &\quad\mbox{if $j_1>0$ and if $j_1=0$, $j_2>0$},\\
\sin\langle j,x\rangle &\quad\mbox{if $j_1<0$ and if $j_1=0$, $j_2<0$},
\end{aligned}
\right.
\end{equation}
where $j^\bot=(-j_2,j_1)$. Let us fix numbers $\{b_j,j\in\Z_*^2\}$ such that 
$$
\BBBB_1<\infty, 
$$
where for $k\ge0$ we denote 
\begin{equation} \label{1.4}
\BBBB_k=\sum_{j\in\Z_*^2}b_j^2|j|^{2k}\le\infty,
\end{equation}
and define 
\begin{equation} \label{1.5}
\eta(t,x)=\frac{\p}{\p t}\zeta(t,x), \quad 
\zeta(t,x)=\sum_{j\in\Z_*^2}b_j\beta_j(t)e_j(x),
\end{equation}
where $\{\beta_j,j\in\Z_*^2\}$ is a family  of independent standard Brownian motions. Then $\eta(t,x) = \sum b_j \eta_j (t) e_j(x)$, where $\{\eta_j =\dot\beta_j, j\in \Z^2_*\}$, are standard independent {\it white noises}. It follows from the Doob--Kolmogorov inequality (see Theorem~3.8 in~\cite[Chapter~1]{KS1991}) that, with probability~$1$, the series in~\eqref{1.5} converges in~$H^1$ uniformly in $t\in[0,T]$ for any $T<\infty$, so that~$\eta(t)$ is the time derivative of a continuous vector function with range in the space $V=H\cap H^1(\T^2,\R^2)$.

\medskip
\noindent
{\bf Random kicks}. Let~$\{\eta^k\}$ be a sequence of i.i.d.\ random variables (the kicks)  with range in~$V$. Define 
\begin{equation} \label{1.6}
\eta(t,x)=\frac{\p}{\p t}\zeta(t,x), \quad 
\zeta(t,x)=\sum_{k=1}^\infty \eta^k(x)\theta(t-k),
\end{equation}
where $\theta(t)=0$ for $t<0$ and $\theta(t)=1$ for $t\ge0$. Since~$\eta$ has  jumps only at positive integers, the trajectories of~$\eta$ are the time derivative of piecewise continuous functions.

\medskip
\noindent
{\bf Piecewise independent process.}
Let $\{\eta^k\}$ be a sequence of i.i.d.\ random variables in $L^2(J_1,H)$. We define a random process of the form
\begin{equation} \label{1.60}
\eta(t,x)=\sum_{k=1}^\infty I_{[k-1,k)}(t)\eta^k(t-k+1,x), 
\end{equation}
where $I_{[k-1,k)}$ is the indicator function of the interval.

\begin{example}\label{e_red}
Let us take $H$-valued processes $\{\eta^k(t,\cdot), 0\le t\le 1\}$ of the form 
$$
\eta^k(t,x) = \sum_{j\in\Z^2_*} b_j\, \eta^k_j(t) \,e_j(x)\,,\qquad \BBBB_1<\infty\,,
$$
where $\{\eta^k_j: k\ge1, j\in\Z^2_*\}$ are real-valued independent random processes, distributed as a fixed process $\tilde\eta:[0, 1]\to\R$. Taking for~$\tilde\eta$ a random series with respect to the Haar basis of the space $L_2(0,1)$ (see Section~21 in~\cite{lamperti1996}), we arrive at a random process $\eta$ as in~\eqref{1.60} that has the form
\begin{equation} \label{1.61}
\eta(t,x) = \sum_{j\in\Z^2_*} b_j\,\eta_j(t)\,e_j(x)\,.
\end{equation}
Here~$\{\eta_j\}$ are independent random processes, distributed as the process 
\begin{equation} \label{1.62}
\eta_0(t) = c\sum_{l=1}^\infty \xi_l I_{[l-1,l)}(t) +\sum_{N=0}^\infty \sum_{l=0}^\infty c_n \xi^n_l H^n_l(t)\,,
\end{equation}
where $c$ and~$c_n$ are real constants, $\xi_l$ and~$\xi_l^n$ are i.i.d.\ real-valued random variables with a law~$\lambda$, and~$H^n_l$ are the Haar functions:
 \begin{equation*}
 H^n_l (t)=
 \left\{\begin{array}{cl}
0 &\text{if \ $t< l2^{-n}$ or $t\ge (l+1)2^{-n}$},
\\
1&\text{if \ $l2^{-n} \le t <  (l+1/2) 2^{-n}$},
\\
-1 &  \text{if \ $(l+1/2)2^{-n} \le t< (l+1) 2^{-n}$}.
\end{array}\right.
\end{equation*}
Thus, $\eta_0(t)$ is a random wavelet series, and~$\eta(t,x)$ is an $H$-valued process whose expansion in the trigonometric basis~$\{e_j\}$ has independent random wavelet coefficients. Notice that a.s.~the set of discontinuities of~$\eta$ is the family of dyadic numbers and, hence, is dense on the positive half-line. Besides, all trajectories of~$\eta$ are continuous at non-dyadic points and right-continuous everywhere.

If $c=1$, $c_n= 2^{n/2}$, and~$\lambda$ is the centred normal law with unit variance, then~$\eta_0(t)$ is  white noise  (see \cite{lamperti1996}). If $c_n\ll 2^{n/2}$, then~$\eta_0(t)$ is a {\it red noise\/}. In particular, if  $|c_n| \le Cn^{-q}$ for some $q>1$, and $\lambda$ has a bounded support,  then the red noise~$\eta_0(t)$ is bounded uniformly in $t$ and~$\omega$.
\end{example}

\medskip
For reasons of space, we shall usually state the  results for the case of spatially regular white noise. However, suitable reformulations of most   of them remain valid in the two other cases.

\subsection{Markov process and a priori estimates}
\label{s1.2}
The family of solutions, corresponding to the three classes of random forces considered in the foregoing subsection, give rise to  Markov processes in the space~$H$. This allows us to apply to the study of the Navier--Stokes system~\eqref{0.1} with such random forces  well-developed probabilistic methods. We start with Eqs.~\eqref{0.1},~\eqref{1.5} (recalling that always $h\in H$ and $\BBBB_1<\infty$),  and for a random initial data $u_0 = u_0^\omega(x)$, independent from the force $f$, denote by $u(t;u_0)$ a solution of \eqref{0.1}, \eqref{0.2}, \eqref{1.5}. For a non-random $u_0=v\in H$ denote
$$
 P_t(v,\cdot) = \DD(u(t;v))\,.
$$
This is a measure in $H$, depending on $(t,v)$ in a measurable way, and satisfying the Kolmogorov--Chapman relation. So~$P_t$ is the transition function of a Markov process in~$H$. The latter is the Markov process generated by solutions  of Eqs.~\eqref{0.1},  \eqref{1.5}. It defines  the {\it Markov semigroups\/} on functions and on measures by the relations
\begin{align}
\PPPP_t&: C_b(H)\to C_b(H), &\quad (\PPPP_tg)(u)&=\int_HP_t(u,\dd z)g(z),
\label{1.010}\\
\PPPP_t^*&: \PP(H)\to \PP(H), &\quad 
(\PPPP_t^*\mu)(\Gamma)&=\int_HP_t(z,\Gamma)\mu(\dd z). 
\label{1.011}
\end{align}
The two semigroups are instrumental to study the equation since the former defines the evolution of mean values of observables: for  any $g\in C_b(H)$ and any $v\in H$, we have
\begin{equation} \label{1.9}
\E \,g(u(t;v))=(\PPPP_tg)(v), \quad t\ge0.
\end{equation}
On the other hand, the latter defines the evolution of the laws since 
\begin{equation} \label{1.10}
\DD(u(t;u_0))=\PPPP_t^*\mu, \quad t\ge0,
\end{equation}
if $u_0$ is a random variable independent from~$f$, and its law equals~$\mu$.

The white in time structure of the noise  allows  not only to prove the Markovian character of evolution, but also to derive a priori estimates by an application of Ito's formula. The  theorem below summarises some of them. For any integer $k\ge0$, we set
$$
\EE_u(k,t)=t^k\|u(t)\|_k^2+\int_0^ts^k\|u(s)\|_{k+1}^2\dd s, \quad t\ge0. 
$$
In the case $k=0$, we write $\EE_u(t)$. For $k\in\N$ we denote $H^k = H\cap H^k(\T^2,\R^2)$ (so $H^1=V$).

\begin{theorem} \label{t1.4}
Consider Eqs.~\eqref{0.1}, \eqref{1.5}. The following properties hold for any $\nu>0$ and any $H$-valued random variable~$u_0$, independent from~$\zeta$.

\medskip
\noindent{\bf Energy and enstrophy balances.}
If $\E|u_0|_2^2<\infty$, then
\begin{equation} \label{1.12}
\E\,|u(t)|_2^2+2\nu\E\int_0^t|\nabla u(s)|_2^2ds
=\E\,|u_0|_2^2+\BBBB_0 t+2\,\E\int_0^t(u,h)\dd s.
\end{equation}
If, in addition, $\E\,\|u_0\|_1^2<\infty$, then
\begin{equation} \label{1.13}
\E\,\|u(t)\|_1^2+2\nu\E\int_0^t|\Delta u(s)|_2^2ds
=\E\,\|u_0\|_1^2+\BBBB_1 t+2\,\E\int_0^t(\nabla u,\nabla h)\dd s.
\end{equation}

\smallskip
\noindent{\bf Time average.}
There is $\gamma>0$ depending only on~$\{b_j\}$ such that
\begin{equation*}
\IP\left\{\,\sup_{t\ge0}
\bigl(\EE_u(t)-(\BBBB_0 +2\nu^{-1}|h|_2^2)\,t\,\bigr)\ge |u_0|_2^2+\rho\right\} 
\le e^{-\gamma \nu\rho}, \quad \rho>0.
\end{equation*}

\smallskip
\noindent{\bf Exponential moment.}
There is $c>0$ not depending on~$\nu$, $h$ and~$\{b_j\}$ such that, if $\varkappa>0$ and~$u_0$ satisfy the inequalities
\begin{equation*}
\varkappa \sup_{j\ge1}b_j^2\le c,\quad \E\exp\bigl(\varkappa\nu|u_0|_2^2\bigr)<\infty,
\end{equation*}
then, for some number $K=K(\nu,\varkappa,\BBBB_0,h)$, we have
\begin{equation} \label{1.14}
\E\exp\bigl(\varkappa\nu|u(t)|_2^2\bigr)
\le e^{-\varkappa\nu^2 t}\E\exp\bigl(\varkappa\nu|u_0|_2^2\bigr)+K,
\quad t\ge0.
\end{equation}

\smallskip
\noindent{\bf Higher Sobolev norms.}
Suppose that $h\in  H^k$ and $\BBBB_k<\infty$ for some integer~$k\ge1$. Then, for any $m\ge1$ and $T\ge1$, there is $C(k,m,T)>0$ such that
\begin{equation} \label{1.15}
\E\sup_{0\le t\le T}\EE_u(k,t)^m
\le C(k,m,T)\bigl(1+\nu^{-m(7k+2)}\bigl(\E\,|u_0|_2^{4m(k+1)}+1\bigr)\bigr). 
\end{equation}
\end{theorem}

A straightforward consequence of the energy balance~\eqref{1.12} and Gronwall's inequality is the exponentially fast stabilisation of the~$L^2$ norms of solutions:
\begin{equation*}
\E\,|u(t)|_2^2\le e^{-\nu t}\E\,|u_0|_2^2+\nu^{-1}\BBBB_0
+\nu^{-2}|h|_2^2,\quad t\ge0. 
\end{equation*}
Combining this with~\eqref{1.14} and~\eqref{1.15}, we conclude that, if all moments of 
$|u_0|_2^2$ are finite,  then
\begin{equation} \label{1.18}
\E\sup_{s\le t\le s+T}\EE_u(k,t)^m\le C'(k,m,T)
\quad\mbox{for all $s\ge0$},\; m\in\N\,.
\end{equation}
A proof of all these results, as well as of their counterparts for random kick--forces,  can be found in Chapter~2 of~\cite{KS-book}.
\medskip

The Markov process defined by solutions of Eqs.~\eqref{0.1},~\eqref{1.5} is time-homo\-ge\-neous: the law at time~$t_2$ of a solution~$u(t)$ which takes a prescribed deterministic value~$v$ at $t=t_1<t_2$ depends only on $v$ and $t_2-t_1$. Solutions of the kick-forced equation  \eqref{0.1},~\eqref{1.6} define an inhomogeneous Markov process, but its restriction to integer values of time $t\in\Z$ is a homogeneous Markov chain, and when studying Eqs.~\eqref{0.1},~\eqref{1.6} we usually restrict ourselves to  integer $t$'s,  see in \cite{KS-book}. Finally, solutions~$u(t)$  of Eqs.~\eqref{0.1}, \eqref{1.60} do not define a Markov process, but their restrictions to $t\in\Z$  form  a homogeneous Markov chain, which is the subject of our study when dealing with that equation, see \cite{shirikyan-asens2015,KNS-2017}.

\section{Mixing}
\label{s2} 
In the previous section, we discussed the existence, uniqueness, and regularity of the flow for the Navier--Stokes system subject to an external random force. Our next goal is to study its  large-time asymptotics.  As before, we shall mostly concentrate on the case of  spatially regular white noise. 

\subsection{Existence of a stationary measure}
\label{s2.1}
We consider the Navier--Stokes system~\eqref{0.1} with the right-hand side of the form~\eqref{1.2}, where $h\in H$ is a deterministic function and~$\eta$ has the form~\eqref{1.5}. Since  $\BBBB_1<\infty$, then $\zeta$ is a continuous functions of time with range in~$V$. 

\begin{definition} \label{d2.1}
A measure $\mu\in\PP(H)$ is said to be {\it stationary\/} for the Navier--Stokes system if $\PPPP_t^*\mu=\mu$ for all $t\ge0$. 
\end{definition}

By \eqref{1.9},  \eqref{1.10}, a measure $\mu\in\PP(H)$ is stationary if and only if there is an $H$-valued random variable~$u_0$ independent from~$\zeta$ such that $\DD(u_0)=\mu$, and one of the following equivalent properties is satisfied for the corresponding solution~$u(t;u_0)$:

\smallskip
{\bf (a)} for all $g\in C_b(H)$ and $t\ge0$, we have $\E\,g(u(t))=\E\,g(u_0)$;

\smallskip
{\bf (b)} the measure~$\DD(u(t))$ coincides with~$\mu$ for all $t\ge0$. 

\smallskip
\noindent
A solution $u(t)$ of \eqref{0.1} as in (a) (or (b)\,) is called a {\it stationary solution}. 

\smallskip
The existence of a stationary measure for the 2d Navier--Stokes system can be established with the help of the Borolyugov--Krylov argument, even though the first works dealing with both 2d and 3d cases used a different  approach; see~\cite{VF1988}. We refer the reader to the article~\cite{flandoli-1994} a for self-contained and  simple proof of the existence of a stationary measure. In addition,  Theorem~\ref{t1.4} jointly with Fatou's lemma imply  a priori estimates for any stationary measure of~\eqref{0.1}. A detailed proof of the following result can be found in~\cite[Chapter~2]{KS-book}. 

\begin{theorem} \label{t2.2} 
The Navier--Stokes system \eqref{0.1}, \eqref{1.5} has at least one stationary measure $\mu_\nu$. It is supported by the space~$H^2$ and satisfies the {\rm energy and enstrophy balance relations}
\begin{align}
\nu\int_H |\nabla u|_2^2\mu_\nu(\dd u) 
&=\frac12 \BBBB_0+\int_H(h,u)\mu_\nu(\dd u), \label{1.19}\\
\nu\int_H|\Delta u|_2^2\mu_\nu(\dd u) 
&=\frac12 \BBBB_1+\int_H(\nabla h,\nabla u)\mu_\nu(\dd u). \label{1.20}
\end{align} 
 If, in addition,  $h\in C^\infty$ and~$\BBBB_k<\infty$ for all  $k\ge0$, then every stationary measure $\mu_\nu\in\PP(H)$ is concentrated on infinitely smooth functions\,\footnote{This means that the $\mu_\nu$-measure of the space of $C^\infty$ functions is equal to~$1$.}, and  there are positive numbers $\varkappa$, $C$, and~$C_{km}$ not depending on~$\nu$ such that, for any integers~$m\ge1$ and~$k\ge2$, we have 
\begin{align}
\int_H\exp\bigl(\varkappa\nu\|u\|_1^2\bigr)\mu_\nu(\dd u)&\le C,\label{1.21}\\
\nu^{m(7k+2)}\int_H\|u\|_k^{2m}\mu_\nu(\dd u)&\le C_{km}.\label{1.22}
\end{align}
\end{theorem}

Besides,  for stationary solutions $u(t,x)$ of Eq.~\eqref{0.1}, estimate~\eqref{1.15} implies  uniform in $s\ge0$ bounds of the form  
$$
\E\,\sup_{s\le t\le s+T}\|u(t)\|_k^{2m}\le C(k,m,T)\,\nu^{-m(7k+2)}\,;
$$
see Corollary~2.4.13 in~\cite{KS-book}. 

\subsection{Uniqueness and exponential stability}
\label{s2.2}
In contrast to the existence of a stationary measure (which is established by rather soft tools), its uniqueness is a deep result that was proved thanks to the contribution of various research groups. It was first established in the case of spatially irregular white noise by Flandoli and Maslowski~\cite{FM-1995} and then extended to various types of regular noises in~\cite{KS-cmp2000} and next in \cite{EMS-2001,BKL-2002,KS-cmp2001,kuksin-ams2002,mattingly-2002,KS-jmpa2002,KS-prse2003,shirikyan-rjmp2005,odasso-2008} (see Chapter~3 in~\cite{KS-book} for more references). The following theorem summarises those results in the case of spatially regular white noise.

\begin{theorem} \label{t2.3}
If the random process $\zeta$ in \eqref{1.5} satisfies 
\begin{equation} \label{2.5}
b_j\ne0\quad\mbox{for all $j\in\Z_*^2$}\,,
\end{equation}
then the problem \eqref{0.1}, \eqref{1.5}  has a unique stationary distribution $\mu_\nu\in\PP(H)$. This measure possesses the following properties.
\begin{description}
\item[Exponential mixing.]
There are positive numbers~$\gamma_\nu$ and~$\varkappa_\nu$ such that, for any locally H\"older-continuous function $g:V\to \R$ with at most exponential growth at infinity and any $H$-valued random variable~$u_0$ independent from~$\zeta$, we have
\begin{equation} \label{2.6}
\bigl|\E\,g(u(t;u_0))-\langle g,\mu_\nu\rangle\bigr|
\le C(\nu,g)e^{-\gamma_\nu t}\,\E\,e^{\varkappa_\nu|u_0|_2^2}, \quad t\ge1,
\end{equation}
where  $C(\nu,g)$  depends on~$\nu$ and a specific norm of~$g$, but not on~$u_0$.
\item[Convergence for observables.]
If, in addition,  $h\in H^k$ and $\BBBB_k<\infty$ for all~$k$, then a similar convergence holds for any H\"older-continuous function~$g$ that is defined on a Sobolev space~$H^s$ of any finite order and has at most polynomial growth at infinity.
\item[Space homogeneity.]
If, in addition to~\eqref{2.5}, $b_s\equiv b_{-s}$, then the measure $\mu_\nu$ is space homogeneous (i.e. the space translations  $H\ni u(x) \mapsto u(x+y)$, $y\in\T^2$, do not change it). 
\end{description}
\end{theorem}

The first assertion of the theorem implies that the Markov process in $H$ which  we discuss  is {\it exponentially mixing in the Lipschitz--dual distance in the space~$H$\/}, i.e.~for each measure $\rho \in\PP(H)$ with a finite second exponential moment we have
\begin{equation} \label{mix}
\|\PPPP_t^*\rho-\mu_\nu\|^*_{\text{Lip}(H)}
 \le C_\nu e^{-\gamma'_\nu t}\quad \text{for}\quad t\ge0\,
\end{equation}
with some $C_\nu, \gamma'_\nu>0$, where 
  \begin{equation} \label{LD}
  \|\rho_1 - \rho_2 \|^*_{\text{Lip}(H)} = \sup \langle \rho_1 - \rho_2 , f\rangle\,,\quad \langle \rho, f\rangle = \int_H f(v)\,\rho(dv)\,,
\end{equation}  
and  the supremum  is taken oven all Lipschitz functions~$f$  on~$H$ whose norm and Lipschitz constant are bounded  by one. If~\eqref{mix} holds, we also  say that the stationary measure $\mu_\nu$ and Eq.~\eqref{0.1} are exponentially mixing (in $H$). 
%If the l.h.s.\ of~\eqref{mix}  simply converges to zero when time goes to infinity, we say that the equation,  Markov process and measure~$\mu_\nu$ are {\it mixing} (in~$H$). 

Inequality~\eqref{2.6} expresses the property of convergence of the ensemble average of an observable~$g$ to its mean value with respect to the stationary measure. It applies to various physically relevant quantities, such as the energy~$\tfrac12|u|_2^2$, the enstrophy~$\tfrac12|\nabla u|_2^2=\tfrac12|\rot u|_2^2$, and the correlation tensors~$u_i(x)u_j(y)$, where $x,y\in\T^2$ are arbitrary points. 

Similar results hold for solutions of Eqs.~\eqref{0.1}, \eqref{1.6} if in~\eqref{2.6} we replace $t\ge1$ by $t\in\N$. 

In Theorem~\ref{t2.3}, the rate of convergence depends on the viscosity~$\nu$: when the latter decreases, the attractor of the unperturbed problem becomes larger and more chaotic, and it seems to be a very complicated 
task to establish a uniform convergence to the limiting measure. On the other hand, it is not difficult to prove that, when $\nu>0$ is fixed, the non-degeneracy condition~\eqref{2.5} can be relaxed, requiring only that the noise should act directly on the (finitely many{\footnote{The fact that dissipative parabolic-type PDEs have only finitely many determining modes goes back to the paper~\cite{FP-1967} and implies, in particular, that the global attractor is finite-dimensional.}) determining modes of the dynamics. A challenging problem, important, in particular, for numerical simulations, is to prove that the mixing property remains true under a weaker hypothesis on the noise, allowing for the noise's localisation in a part of either the physical or the Fourier spaces, so that the determining modes of the dynamics do not necessarily belong to  the region of the phase space affected by the noise. Propagation of the randomness then may take  place due to the ``mixing'' properties of the (deterministic) Navier--Stokes flow. Uniqueness and mixing of the random flow in this situation are mostly established  in the case when the deterministic component~$h$ of the random force is zero. 

We begin with the case when the random force is localised in the Fourier space. In~\cite{HM-2006,HM-2011} Hairer and Mattingly obtained the following result:
  
  \begin{theorem}\label{t.HM}
  Let the random force $f$ have the form \eqref{1.2}, \eqref{1.5} with $h=0$ and with~$\eta$  satisfying 
\begin{equation} \label{2.7}
b_j\ne0\quad\mbox{if and only if  $j\in\II$},
\end{equation}
where $\II$ is a finite subset of $\Z^2_*$ such that any vector in~$\Z^2$ can be represented as an integer linear combination  of the  elements of~$\II$, and~$\II$ contains at least two vectors of different length. Then Eq.~\eqref{0.1} is exponentially mixing in~$H$.
  \end{theorem}

 A key ingredient of the proof is an infinite-dimensional version of the Malliavin calculus, which uses the white noise structure of the random perturbation. Note that the result above does not imply the assertion of Theorem~\ref{t2.3} since now the set of modes $j\in \Z^2_*$, excited by the random force, must be finite. 
 
 The recent paper~\cite{KNS-2017} deals with the case when the noise is a piecewise independent random process of the form~\eqref{1.60}. The main result of  \cite{KNS-2017}  is an abstract theorem, establishing the mixing property for a large class of nonlinear PDE, perturbed by bounded random forces of the form \eqref{1.2}, \eqref{1.60}; its proof relies on the method of optimal control and a variant of the Nash--Moser scheme. In particular, the theorem in \cite{KNS-2017}  applies to the Navier--Stokes system, perturbed by a bounded red noise as in Example~\ref{e_red}:
 
  \begin{theorem}\label{t.KSN}
  Let the random force~$f$ have the form \eqref{1.2}, in which $h=0$ and~$\eta$ is given \eqref{1.61}, \eqref{1.62}, with the set of coefficients~$b_j$ satisfying~\eqref{2.7}. Assume that $|c_n|\le Cn^{-q}$ for all~$n\ge1$ with some $q>1$, and that  the law of the random variables~$\xi_l$ and~$\xi^n_l$  has the form $\lambda= p(r)\,\dd r$ with $p\in C_0^1(-1, 1)$, $p(0)\ne0$.  Then Eq.~\eqref{0.1} is exponentially mixing in~$H$.
  \end{theorem}

The situation in which the random perturbation is localised in the physical space was studied in~\cite{shirikyan-asens2015} (see also~\cite{shirikyan-2017} for the case of a boundary noise). To formulate the corresponding result, we fix a non-empty open set $Q\subset \R\times\T^2$ whose closure is contained in $(0,1)\times\T^2$ and denote by~$\{\varphi_j\} \subset H^1(Q,\R^2)$ an orthonormal basis in~$L^2(Q,\R^2)$. Considering again the random force~\eqref{1.2}, we assume that $h\equiv0$, and~$\eta$ is a piecewise independent random process of the form~\eqref{1.60}, with 
$$
\eta^k(t,x)=\sum_{j=1}^\infty b_j\xi_{jk}\psi_j(t,x).
$$
Here $\psi_j=\chi\varphi_j$, where $\chi\in C_0^\infty(Q)$ is a nonzero function, $\{b_j\}$ are real numbers such that $\sum_jb_j\|\psi_j\|_1<\infty$, and $\xi_{jk}$ are independent random variables whose laws have have the form $\lambda_j=p_j(r)\,\dd r$, where $p_j\in C_0^1(-1, 1)$ and $p_j(0)\ne0$. Thus, the random force entering the right-hand side of~\eqref{0.1} is bounded and space-time localised in~$Q$. The following theorem is the main result of~\cite{shirikyan-asens2015}. 

\begin{theorem} \label{t2.6}
	In addition to the above hypotheses, assume that $b_j\ne 0$ for all~$j\ge 1$. Then, for any $\nu>0$, Eq.~\eqref{0.1} is has a unique stationary measure, which is exponentially mixing in~$H$. 
\end{theorem}

\section{Consequences of mixing}
\label{s3}
The results of the previous section concern the evolution of the mean values of observables and the laws of solutions under the stochastic   Navier--Stokes flow. This section is devoted to studying the typical behaviour of individual trajectories. We will assume that the random force in equation \eqref{0.1} is such that the equation is exponentially mixing, either for $t>0$, or for $t\in\N$. That is, either the assumptions of Theorems~\ref{t2.3}, \ref{t.HM}, \ref{t.KSN} or~\ref{t2.6} hold, or~$f$ is a kick-force of the form~\eqref{1.2}, \eqref{1.6}, where \eqref{2.5} holds. We state the results for the case when Theorem~\ref{t2.3} applies. Situation with the other cases is very similar; e.g., see~\cite{KS-book} for the case of the kick-forced equations.

\subsection{Ergodic theorems}
\label{s3.1}
We consider the Navier--Stokes system~\eqref{0.1}, \eqref{1.2} in which $h\in H$ is a fixed function and~$\eta$ is given by~\eqref{1.5}.  Let us denote\footnote{To avoid unimportant complications, we do not give an exact definition of the space~$\HH$, referring the reader to Section~4.1 in~\cite{KS-book}.} by~$\HH$ the space of locally H\"older-continuous functions $g:V\to\R$ with at most exponential growth at infinity. The following result established in~\cite{kuksin-rmp2002,shirikyan-ptrf2006} (see also Section~4.1.1 in~\cite{KS-book}) shows that the time average of a large class of observables converges to their mean value with respect to the stationary measure.

\begin{theorem}[Strong law of large numbers] \label{t3.1}
Under the hypotheses of Theorem~\ref{t2.3}, for any $\gamma\in[0,1/2)$, any non-random $u_0\in H$ and $g\in \HH$, with probability~$1$ we have 
\begin{equation} \label{3.1}
\lim_{t\to\infty} t^\gamma
\biggl(\frac1t\int_0^t g(u(s;u_0))\,\dd s-\langle g,\mu_\nu\rangle \biggr)=0\,.
\end{equation}
\end{theorem}

Convergence~\eqref{3.1} remains true for random initial functions that are independent from~$\eta$ and have a finite exponential moment. Moreover, some further analysis shows that the  law of iterated logarithm (LIL) is also valid. In particular,  the number~$\gamma$ in~\eqref{3.1} characterizing the rate of convergence to the mean value cannot be taken to be equal to~$1/2$; see Section~4.1.2 in~\cite{KS-book}. 

We now turn to the central limit theorem (CLT). To this end, given an observable $g\in\HH$ with zero mean value with respect to~$\mu_\nu$, we denote 
$$
\sigma_g^2=2\int_H\int_0^\infty(\PPPP_tg)(v)\dd t \,g(v)\mu_\nu(\dd v). 
$$
In view of inequality~\eqref{2.6} and the assumption that $\langle g,\mu_\nu\rangle=0$, the function~$\PPPP_tg$ decays exponentially to zero as $t\to\infty$, and it is not difficult to prove that, under the hypotheses of Theorem~\ref{t2.3}, the number~$\sigma_g^2$ is well defined and positive for any non-constant~$g$; see Proposition~4.1.4 in~\cite{KS-book}. A proof of the following result can be found in~\cite{kuksin-rmp2002,shirikyan-ptrf2006} (see also Section~4.1.3 in~\cite{KS-book}).

\begin{theorem}[Central limit theorem]  \label{t3.2}
Under the hypotheses of Theorem~\ref{t2.3}, for any non-random
$u_0\in H$ and any non-constant $g\in\HH$ satisfying $\langle g,\mu_\nu\rangle=0$, we have 
\begin{equation} \label{3.2}
\DD\biggl(\frac{1}{\sqrt{t}}\int_0^tg(u(s;u_0))\,\dd s\biggr)\rightharpoonup\NN_{\sigma_g}\quad\mbox{as $t\to\infty$},
\end{equation}
where $\NN_\sigma$ stands for the centred normal law on~$\R$ with a variance~$\sigma^2>0$, and~$\rightharpoonup$ stands for the weak convergence of measures.
\end{theorem}

Let us emphasise that~$\sigma_g$ is expressed in terms of the stationary measure~$\mu_\nu$ and does not depend on~$u_0$. Furthermore, convergence~\eqref{3.2} is equivalent to the relation
\begin{equation} \label{3.3}
\lim_{t\to\infty}\IP\biggl\{\frac{1}{\sqrt{t}}\int_0^tg(u(s))\,\dd s\in\Gamma\biggr\}
=\NN_{\sigma_g}(\Gamma), 
\end{equation}
where $\Gamma\subset\R$ is an arbitrary Borel set whose boundary has zero Lebesgue measure. 

\subsection{Random attractors}
\label{s3.2}
Another important object characterising the large-time behaviour of trajectories, is the random attractor. There are many definitions of this object, and in the context of random dynamical systems most of them deal with the concept of pullback attraction.  The meaning of the latter is that, for some fixed observation time, the distance between trajectories of the system and the attractor decreases when the moment of beginning of the observation goes to~$-\infty$. In this section, we discuss a concept of attractor that possesses an  attraction property forward in time and is closely related to the unique stationary measure constructed in Theorem~\ref{t2.3}. To simplify notation, we fix the viscosity~$\nu>0$ and do not follow the dependence of various objects on it. Moreover, we assume that the Brownian motions~$\{\beta_j\}$ entering~\eqref{1.5} are two-sided\footnote{That is, they are defined for all $t\in\R$ and vanish at $t=0$.} and denote by~$(\Omega,\FF,\FF_t,\IP)$ the canonical filtered probability space associated with the process~$\zeta$; see Section~2.4 in~\cite{KS1991}. In particular, the measurable space $(\Omega, \FF)$ may  be chosen to coincide with the Fr\'echet space $C(\R,V)$ of continuous functions from~$\R$ to~$V$, with the topology of uniform convergence on bounded intervals  and the corresponding Borel sigma-algebra. 

Let us denote by $\varphi_t(\omega):H\to H$ the random flow generated by the Navier--Stokes system with spatially regular white noise~\eqref{1.5}. Thus, for any initial function $u_0\in H$, the solution $u(t;u_0)$ 
of~\eqref{0.1}, \eqref{0.2}, \eqref{1.5} is given by $\varphi_t(\omega)u_0$. The family~$\{\varphi_t(\omega), \omega\in\Omega,t\ge0\}$ possesses the perfect co-cycle property. Namely, let $\theta_t:\Omega\to\Omega$ be the shift operator taking~$\omega(\cdot)$ to~$\omega(\cdot+t)$. Then, there is a set of full measure $\Omega_0\in\FF$ such that, for any $\omega\in\Omega_0$, we have 
\begin{equation*} 
\varphi_{t+s}(\omega)u_0=\varphi_t(\theta_{s}(\omega))\varphi_s(\omega)u_0
\quad\mbox{for all $t,s\ge0$, $u_0\in H$}. 
\end{equation*}
The following result is established in~\cite{ledrappier-1986,lejan-1987,crauel-1991} (see also Theorem~4.2.9 in~\cite{KS-book}) in the context of general random dynamical systems. 

\begin{theorem} \label{t3.3}
Under the hypotheses of Theorem~\ref{t2.3}, for any sequence~$\{t_k\}$ going to~$+\infty$ the limit 
\begin{equation} \label{3.4}
\mu_\omega=\lim_{k\to\infty}\varphi_{t_k}(\theta_{-t_k}(\omega))_*\mu
\end{equation}
exists in the weak topology of~$\PP(\Omega)$ for almost every $\omega\in\Omega$. Moreover, the following properties hold:
\begin{description}
\item[Uniqueness.]
If~$\{t_k'\}$ is another sequence going to~$+\infty$ and~$\{\mu_\omega'\}$ is the corresponding limit, then $\mu_\omega=\mu_\omega'$ almost everywhere.
\item[Reconstruction.]
For any $\Gamma\in\BB(H)$, the mapping $\omega\mapsto\mu_\omega(\Gamma)$ is measurable, and~$\mu$ can be reconstructed by the formula $\mu(\Gamma)=\E\mu_\cdot(\Gamma)$. 
\end{description}
\end{theorem}

We now describe a random  attractor associated with~$\mu$. To this end, we fix any sequence~$\{t_k\}$ going to~$+\infty$ and denote by~$\Omega_0$ the set of full measure on which limit~\eqref{3.4} exists. The almost sure existence of the limit in~\eqref{3.4} implies that~$\Omega_0$ is $\IP$-invariant under~$\theta_t$; that is, $\IP(\Omega_0\triangle(\theta_t(\Omega_0))=0$ for any $t\in\R$, where~$\triangle$ stands for the symmetric difference of two sets. We define~$\mu_\omega$ by~\eqref{3.4} on the set of full measure~$\Omega_0$ and denote by~$\aA_\omega$ the support of~$\mu_\omega$ for $\omega\in\Omega_0$, while we set $\aA_\omega=\varnothing$ on the complement of~$\Omega_0$. The measurability property of~$\mu_\cdot$ mentioned in Theorem~\ref{t3.3} implies that~$\{\aA_\omega,\omega\in\Omega\}$ is also measurable in the sense that, for any $u\in H$, the function $\omega\mapsto d_H(u,\aA_\omega)$ is measurable. The following result is established in~\cite{KS-faa2004} (see also Section~4.2 in~\cite{KS-book}). 

\begin{theorem} \label{t3.4}
Under the hypotheses of Theorem~\ref{t2.3}, the following properties hold:
\begin{description}
\item[Invariance.]
For any $t\ge0$ and almost every $\omega\in\Omega$, we have $\varphi_t(\omega)\aA_\omega=\aA_{\theta_t(\omega)}$. 
\item[Attraction.]
For any $u\in H$, the functions $\omega\mapsto d_H(\varphi_t(\omega)u,\aA_{\theta_t(\omega)})$ converge to zero in probability as $t\to\infty$. That is, for any $\e>0$, we have
\begin{equation*}
\IP\bigl\{d_H(\varphi_t(\omega)u,\aA_{\theta_t(\omega)})\ge\e\bigr\}\to0
\quad\mbox{as $t\to\infty$}. 
\end{equation*}
\item[Minimality.]
If $\{\aA_\omega',\omega\in\Omega\}$ is another measurable family of closed subsets that satisfies the first two properties, then $\aA_\omega\subset\aA_\omega'$ for almost every $\omega\in\Omega$. 
\end{description} 
\end{theorem}

\subsection{Dependence on parameters and stability}
\label{s3.3}
We now investigate how the laws of trajectories of~\eqref{0.1} vary with parameters. To simplify the presentation, we shall consider only the dependence on the random forcing, assuming that it is a spatially regular white noise. However, similar methods can be used to study the dependence on other parameters, such as the viscosity, as well as the relationship between  stationary measures corresponding to various types of random forcings. 

We thus assume that 
$X$ is a metric space and that the right-hand side in~\eqref{0.1} has the 
form~\eqref{1.2}, \eqref{1.5}, where $h\in H$ is fixed and $b_j$'s  are continuous functions of $a\in X$, satisfying 
\begin{equation*} 
\sup_{a\in X}\sum_{j\in\Z_*^2} b_j(a)^2|j|^2<\infty.
\end{equation*}
A proof of the following result can be found in Section~4.3.1 of~\cite{KS-book}.

\begin{theorem} \label{t3.5}
In addition to the above hypotheses, let
\begin{equation}
b_j(\hat a)\ne0\quad\mbox{for all $j\in\Z_*^2$},
\end{equation}
where $\hat a\in X$ is a fixed point, and let $\{\mu^a\}$ be a family of stationary measures\footnote{Existence of a measure~$\mu^a$ is guaranteed by Theorem~\ref{t2.2}, and Theorem~\ref{t2.3} implies that the measure~$\mu^{\hat a}$ is uniquely determined.} of Eq.~\eqref{0.1} with the right-hand side corresponding to the value~$a\in X$ of the parameter. Then $\mu^a\to\mu^{\hat a}$ as $a\to\hat a$, and for any compact subset $\Lambda\subset\PP(H)$ there exists 
a continuous function $A_\Lambda(\rho)>0$ going to zero with~$\rho$ such that 
$$
\sup_{t\ge0}\bigl\|\PPPP_t^*(a)\lambda_1-\PPPP_t^*(\hat a)\lambda_2\bigr\|_L^*
\le A_\Lambda\bigl(\|\lambda_1-\lambda_2\|_L^*+d_X(a,\hat a)\bigr)
\quad \mbox{for $\lambda_1,\lambda_2\in \Lambda$, $a\in X$,}
$$
where $\PPPP_t^*(a)$ denotes the Markov semigroup for~\eqref{0.1} corresponding to the parameter~$a$. 
\end{theorem}

We emphasise that by this result the law of a solution for \eqref{0.1} depends on the parameters of the random force~$f$ {\it continuously and uniformly in time}; cf.~Theorem~\ref{t6.1} below. 

\section{Large deviations}
\label{s4}
Having discussed the typical behaviour of trajectories, we now turn to a description of probabilities of rare events. Two different asymptotics will be studied: deviations of the time-average of observables from their ensemble average as $t\to\infty$, and deviations from the limiting dynamics in the small noise regime. In the latter setting, we shall only discuss the behaviour of a stationary distribution, since in the case of an additive noise the asymptotics of trajectories with given initial data is a simple result that follows immediately from the large deviations principle (LDP) for Gaussian random variables. We refer the reader to the paper~\cite{CM-2010} and the references therein for this type of results for stochastic PDEs with multiplicative noise. 

\subsection{Donsker--Varadhan type large deviations}
\label{s4.1}
We first describe the general idea. To this end, let us note that, for an arbitrary function $g\in\HH$ (whose mean value is not necessarily zero), we can rewrite the convergence~\eqref{3.3}  in the form
\begin{equation} \label{4.1}
\lim_{t\to\infty}\IP\biggl\{\frac{1}{t}\int_0^tg(u(s))\,\dd s\in \langle g,\mu_\nu\rangle+\frac{\Gamma}{\sqrt{t}}\biggr\}=\NN_{\sigma_g}(\Gamma), 
\end{equation}
Thus, the CLT can be interpreted as a description of the probabilities of small deviations of the time average of observables from their mean value. The goal of the theory of large deviations is to describe the probabilities of order~$1$ deviations (when $\Gamma/\sqrt{t}$ in~\eqref{4.1} is replaced by~$\Gamma$). This type of results were first obtained by Donsker and Varadhan in the case of finite-dimensional diffusion processes~\cite{DV-1975} and later extended to many other situations. In the context of the Navier--Stokes equations, the theory was developed in the case of random kicks~\cite{JNPS-cpam2015,JNPS-2014} (see also the recent papers~\cite{MN-2015, nersesyan-2017} devoted to spatially regular white noise), and we now describe the main achievements.  

Let us consider the Navier--Stokes system~\eqref{0.1}, \eqref{1.2}, where $h\in H$ is a deterministic function, and the random forcing~$\eta$ is given by~\eqref{1.6}. In this case, the trajectories of~\eqref{0.1} have jumps at integer times, and we normalise them to be right-continuous in time. Setting $u_k=u(k)$, we see that the random sequence~$\{u_k\}$ satisfies the relation
\begin{equation} \label{4.2}
u_k=S(u_{k-1})+\eta^k, \quad k\ge1, 
\end{equation}
where $S:H\to H$ denotes the time-$1$ shift along trajectories of Eq.~\eqref{0.1} with $f=h$. Equation~\eqref{4.2} defines a discrete-time Markov process in~$H$, and we use the notation introduced in Section~\ref{s1.2} to denote the related objects, replacing~$t$ with~$k$. We thus write $P_k(v,\Gamma)$, $\PPPP_k$, and~$\PPPP_k^*$. 

To formulate the result on LDP, we shall need some hypotheses on the kicks $\eta^k$. Namely, we shall assume that they satisfy the following condition, in which~$\{e_j\}$ is the trigonometric basis in~$H$ defined by~\eqref{1.3}. 

\begin{description}
\item[Structure of the noise.] 
{\sl The random kicks~$\eta^k$ have the form 
\begin{equation*}
\eta^k(x)=\sum_{j\in\Z_*^2}b_j\xi_{jk}e_j(x),
\end{equation*}
where $b_j$ are some non-zero numbers satisfying $\BBBB_1<\infty$ and $\{\xi_{jk}\}$ are independent random variables whose laws possess $C^1$-smooth positive densities~$\rho_j$ with respect to the Lebesgue such that $\int_\R|\rho_j'(r)|\dd r \le1$ for any~$j$.} 
\end{description}
This condition ensures that the Markov process $\{u_k\}$ associated with~\eqref{0.1} has a unique stationary measure~$\mu_\nu$ for any $\nu>0$, and the CLT holds for any Holder-continuous functional $f:H\to \R$ with at most exponential growth at infinity (cf.\ Theorem~\ref{t2.3} and~\ref{t3.2}).  

We now introduce the {\it occupation measures\/}
\begin{equation*}
\mu_n^\omega=\frac1n \sum_{k=0}^{n-1}\delta_{u_k},
\end{equation*}
where $\delta_v\in\PP(H)$ stands for the Dirac mass at $v\in H$, and~$\{u_k\}$ is a trajectory of~\eqref{4.2}. Let us recall that the space~$\PP(H)$ is endowed with the topology of weak convergence. We shall say that a mapping $I:\PP(H)\to[0,+\infty]$ is a {\it good rate function\/} if its sub-level set $\{\lambda\in\PP(H):I(\lambda)\le c\}$ is compact for any $c\ge0$. The following result is established in Section~2.2 of~\cite{JNPS-2014} (see also~\cite{JNPS-cpam2015} for the case of bounded kicks). 

\begin{theorem}
Suppose that the above hypothesis on the structure of the noise is satisfied. Then the following assertions hold for any $\nu>0$: 
\begin{description}
\item[Pressure.]
For any $g\in C_b(H)$ and any deterministic initial function $u_0\in H$, there is a finite limit
\begin{equation*}
Q(g)=\lim_{n\to\infty}n^{-1}\log\E\,\exp\biggl\{\sum_{k=0}^{n-1}g(u_k)\biggr\}
\end{equation*}
that is independent from~$u_0$. Moreover, $Q$ is a $1$-Lipschitz function satisfying the relation $Q(g+C)=Q(g)+C$ for any $C\in\R$. 
\item[Rate function.]
The Legendre transform $I:\PP(H)\to[0,+\infty]$ of~$Q$ defined by 
\begin{equation*}
I(\lambda)=\sup_{g\in C_b(H)}\bigl(\langle g,\lambda\rangle-Q(g)\bigr)
\end{equation*}
is a convex good rate function that can be represented by the Donsker--Varadhan relation
\begin{equation*}
I(\lambda)=\sup_{g\ge1}\int_H\log\frac{g}{\PPPP_1 g}\,\dd\lambda,
\end{equation*}
where the supremum is taken over all functions $g\in C_b(H)$ minorised by~$1$. 
\item[LDP.]
For any random initial function~$u_0$ independent from~$\{\eta^k\}$ such that $\E \exp(\delta|u_0|_2^2)<\infty$ for some $\delta>0$, and any Borel set $\Gamma\subset\PP(H)$, we have 
\begin{equation*}
-I(\dot\Gamma)
\le\liminf_{n\to\infty}\frac1n\log\IP\{\mu_n\in\Gamma\}
\le\limsup_{n\to\infty}\frac1n\log\IP\{\mu_n\in\Gamma\}
\le-I(\overline\Gamma),
\end{equation*}
where $\dot\Gamma$ and~$\overline\Gamma$ denote the interior and closure of~$\Gamma$, respectively, and~$I(A)$ is the infimum of~$I$ on the set~$A$.
\end{description}
\end{theorem}

Assuming faster decay for the coefficients~$b_j$ and considering the Navier--Stokes system in higher Sobolev spaces,
it is not hard to show that a similar result holds in the case when~$H$ is replaced by~$H^s$.
Next, application of the standard techniques of the theory of large deviations implies  that, for any H\"older-continuous function $g:H^s\to\R$ with moderate growth at infinity, the time average $n^{-1}\sum_{k=0}^{n-1}g(u_k)$ satisfies the LDP with a good rate function $I_g:\R\to\R$ that can be expressed in terms of~$I$ by the relation (cf.~Section~1.3 in~\cite{JNPS-cpam2015})
\begin{equation*}
I_g(r)=\inf\{I(\lambda):\lambda\in\PP(H), \langle g,\lambda\rangle=r\}. 
\end{equation*}

\subsection{Vanishing noise limit}
\label{s4.2}
We now go back to the Navier--Stokes system~\eqref{0.1} with a spatially regular white noise and discuss the behaviour of the unique stationary measure as the stochastic component of the noise goes to zero. Namely, let us assume that the external force in~\eqref{0.1} has the form
\begin{equation} \label{4.4}
f(t,x)=h(x)+\sqrt{\e}\,\eta(t,x),
\end{equation}
where $h\in H$ is a fixed function, $\e>0$ is a small parameter, and~$\eta$ is given by~\eqref{1.5}. We assume that the coefficients~$b_j$  are non-zero (and satisfy the inequality $\BBBB_1<\infty$), so that for any~$\nu>0$ and~$\e>0$ there is a unique stationary measure~$\mu_\nu^\e\in\PP(H)$. We are interested in the behaviour of~$\mu_\nu^\e$ for a fixed~$\nu$, so we drop~$\nu$ from the notation in this section and write simply~$\mu^\e$.  In what follows, we assume that the following hypothesis is satisfied.

\begin{description}
\item[Global asymptotic stability.] 
{\sl The flow of the unperturbed Navier--Stokes system~\eqref{0.1}, corresponding to $f=h$, has a unique fixed point~$\hat u\in H$, which is globally asymptotically stable in the sense that any other trajectory converges to it as $t\to+\infty$.}
\end{description}
Notice that this condition is satisfied with~$\hat u=0$ if the deterministic part of the force~\eqref{1.2} vanishes. A simple argument based on the uniqueness of a stationary distribution for the limiting equation implies  that $\{\mu^\e\}$ converges weakly to the Dirac mass concentrated at~$\hat u$. In fact, much more detailed information  about that convergence is available.  Namely, let us denote by $S_t(u_0,f)$ the solution of~\eqref{0.1}, \eqref{0.2} and introduce a {\it quasi-potential\/} by the relation 
$$
\VV(v)=\lim_{r\to0}\inf
\biggl\{\frac12\int_0^T\|f(s)\|_b^2\dd s: T>0, f\in L^2(J_T,H), |S_T(\hat u,f)-v|_2\le r\biggr\},
$$
where the infimum is taken over all~$T$ and~$f$ for which the inequality holds,  and we set $\|g\|_b^2=\sum_j b_j^{-2}(g,e_j)^2$. A proof of the following result can be found in~\cite{martirosyan-2016} (see also~\cite{BC-2017} for the case of spatially irregular noise). 

\begin{theorem} \label{t4.2}
Suppose that global asymptotic stability for the limiting dynamics holds, and the coefficients~$\{b_j\}$ entering~\eqref{1.5} are 
non-zero. Then the function $\VV:H\to[0,+\infty]$ has compact
 sub-level sets in~$H$, vanishes only at the point~$\hat u$, and controls the LDP for the family~$\{\mu^\e\}$; that is, for any $\Gamma\in\BB(H)$, we have 
\begin{equation*}
-\inf_{v\in\dot\Gamma}\VV(v)
\le \liminf_{\e\to0^+}\e\log\mu^\e(\Gamma)
\le \limsup_{\e\to0^+}\e\log\mu^\e(\Gamma)
\le -\inf_{v\in\overline\Gamma}\VV(v).
\end{equation*}
\end{theorem}
The facts that~$\VV$ vanishes only at~$\hat u$ and has compact sub-level sets imply that $\mu^\e(H\setminus B)\sim e^{-c(B)/\e}$ as $\e\to0$, where $B\subset H$ is an arbitrary ball around~$\hat u$, and $c(B)>0$ is a number. Hence, the LDP gives an estimate for the rate of concentration of~$\mu^\e$ around~$\hat u$. Let us also note that, in the case when the limiting
 dynamics is not globally asymptotically stable, it is  still  possible to 
 prove that the family~$\{\mu^\e\}$ is exponentially tight, but the validity of the LDP is not known to hold (see Open problem~\ref{op3} in Section~\ref{s7}). However,  more detailed information on the limiting dynamics would be sufficient to get the LDP. For instance, this is the case when there are finitely many stationary points, and the unperturbed dynamics possesses a global Lyapunov function. We refer the reader to~\cite{martirosyan-2017} for detail.

\section{Inviscid limit}
\label{s5}
If the random force $f$ in Eq.~\eqref{0.1} has the form~\eqref{4.4} with $h=0$ and  with $\eta$  as in~\eqref{1.5}, where  $\nu>0$  is fixed and $\e\to0$, then the corresponding stationary measure~$\mu_\nu^\e$ converges to the delta-measure at $0\in H$, and the results of Section~\ref{s4.2}  describe this convergence in more detail. Now assume that~$\e$ and~$\nu$ both  go to zero in such a way that  $\e=\nu^a$ with some  $a>0$. Applying to $\mu^{\nu^a}_\nu$  relation \eqref{1.19} with $h=0$ and   $b_j := \sqrt\e\, b_j$,  $j\in\Z^2_*$,   we get
 $$
 \int_H |\nabla u|_2^2\, \mu^{\nu^a}_\nu(du) = \frac{\BBBB_0}{2}\,\nu^{a-1}. 
 $$
From this we conclude that the measure $\mu^{\nu^a}_\nu$ may have a non-trivial limit as $\nu\to0$ only if $a=1$. Then 
 Eq.~\eqref{0.1}  becomes 
\begin{equation} \label{5.0}
\p_tu+\langle u,\nabla \rangle u-\nu\Delta u+\nabla p=\sqrt\nu\,\eta(t,x),
\quad\diver u=0\,.
\end{equation}
Assume that \eqref{2.5} holds. Then a stationary measure $\mu^{\nu}_\nu$ of \eqref{5.0} is unique. For short, we re-denote it as~$\mu^\nu$.\footnote{Since now $f$ has the form \eqref{1.2} with $h=0$, then  the uniqueness of the  stationary measure also follows from~\eqref{2.7}, but we need~\eqref{2.5} for the  validity of some results in this section.}

\subsection{Properties of $\mu^\nu$, independent from $\nu$.}
In this subsection, we discuss a number of properties of the stationary measures~$\mu^\nu$ which hold uniformly in~$\nu$. They depend only on the quantities~$\BBBB_0$ and~$\BBBB_1$ and indicate certain  universal properties of the statistical equilibria of Eq.~\eqref{5.0}. 

Relations~\eqref{1.19} and~\eqref{1.20} with $h=0$ and $\BBBB_0=:\nu\BBBB_0$, $\BBBB_1=:\nu\BBBB_1$ imply that, uniformly in~$\nu$, 
\begin{equation} \label{5.1}
\Emn |\nabla u|_2^2 = \frac12 \BBBB_0, \quad   
\Emn |\Delta u|_2^2 = \frac12 \BBBB_1\,.
\end{equation}
Since $|\nabla u|_2^2 \le | u|_2 | \Delta u|_2$, then $\Emn|\nabla u|_2^2 \le (\Emn|u|_2^2)^{1/2} (\Emn|\Delta u|_2^2)^{1/2}$. It follows that $\BBBB_0^2/2\BBBB_1 \le \E |u^\nu|_2^2  \le \frac12 \BBBB_0$,  so that the averaged kinetic energy is bounded below and above. Moreover, uniformly in~$\nu$, the measures $\mu^\nu$ satisfy \eqref{1.21}  with $\nu:=1$  (this follows immediately from the proof in~\cite{KS-book}). Consider a stationary solution $u^\nu(t,x)$, corresponding to~$\mu^\nu$. Then $\E\,|u^\nu|_2^2 = \Emn |u|_2^2 $, and the  Reynolds number of~$u^\nu$ is
$$
\mathrm{Re}_\nu=\frac{ [u^\nu] [x]}{\nu}
=\frac{\big(\E\,|u^\nu|_2^2\big) ^{1/2} \cdot 1}{\nu}\sim\nu^{-1}\,
$$
($[\,\cdot\,]$ stands for the characteristic size of a variable), while the averaged kinetic energy $\tfrac12 (\E\,|u^\nu|_2^2)$ is of order one. So when $\nu\to0$ the solutions~$u^\nu$ describe  space-periodic stationary 2d turbulence.  

Assume that $b_s\equiv b_{-s}$, so  the measures~$\mu^\nu$ are space-homogeneous, and that $b_s$ decay sufficiently fast when $|s|\to\infty$. In this case, as it is shown in  \cite{KP-jsp2005, KS-book},  the measures~$\mu^\nu$ possess additional properties. Namely, let~$g(r)$ be any continuous function, having at most a polynomial growth at infinity. Then, denoting $v = \rot  u$, we have the following {\it balance relation}, valid for all~$\nu>0$:
\begin{equation} \label{5.4}
\Emn \!g(v(t,x))|\nabla v(t,x)|^2=\frac12 (2\pi)^{-2} \BBBB_1 \Emn\! g(v(t,x))
\end{equation}
(by the homogeneity the relation does not depend on $x$). Since $|\Delta u|_2^2=|\nabla v|_2^2$, then relations~\eqref{5.1} and the translational  invariance of~$\mu^\nu$ imply that 
$$
\E^\mu |\nabla v(t,x)|^2 = \frac12 (2\pi)^{-2}\BBBB_1. 
$$
So~\eqref{5.4} means that the random variables~$|\nabla v(t,x)|^2$ and~$g(v(t,x))$ are uncorrelated, for any continuous function  $g$ as above and any $(t,x)$. 

The balance relations~\eqref{5.4}} admit a surprising reformulation. For any $\tau\in\R$ denote by $\Gamma_\tau(\omega)$ the random curve $\{x\in\T^2: v^\omega(t,x) =\tau\}$ (it is well defined for a.a.~$\tau$  and~$\omega$, if $b_s$ decays fast enough). Then
\begin{equation} \label{5.5}
\Emn \int_{\Gamma_\tau(\omega)} |\nabla v^\omega|\,\dd\ell
=\frac12\, (2\pi)^{-1} \BBBB_1 \Emn  \int_{\Gamma_\tau(\omega)} |\nabla v^\omega|^{-1}  \,\dd\ell\,, \quad \text{ for a.a.}\;\; \tau\,,
\end{equation}
where $\dd \ell$ is the length element on ${\Gamma_\tau(\omega)}$, and the existence of the integrals in the l.h.s.~and the r.h.s.~for a.a.~$\tau$ is  a part of the assertion.  This is the  {\it co-area form of the  balance relations}. 
Besides, relations \eqref{5.4} imply the following point-wise exponential estimates
\begin{equation} \label{5.6}
\Emn\Big( e^{\sigma|v(t,x)|} +e^{\sigma|u(t,x)|}  + e^{\sigma|\nabla u(t,x)|^{1/2}}\Big) \le K\quad \forall\, x\,,
\end{equation}
valid
uniformly in $\nu$, where the positive constants $\sigma$ and $K$ depend only on the first few numbers~$\BBBB_j$; see~\cite{KS-book}.

\subsection{Inviscid limit}
Since estimates \eqref{5.1} hold uniformly in $\nu$, the family of measures $\{\mu_\nu, 0<\nu\le1\}$ is tight in~$H^{2-\epsilon}$ and, by Prokhorov's theorem, relatively compact in the space~$\PP(H^{2-\epsilon})$, for any $\epsilon>0$.  So any sequence of measures $\{\mu^{\nu'_j}, \nu'_j\to0\}$, contains a weakly converging subsequence:
\begin{equation} \label{5.7}
\mu^{\nu_j} \to \mu \quad \text{weakly in } \quad \PP (H^{2-\epsilon})\,.
\end{equation}
Relations \eqref{5.1} immediately imply that 
\begin{equation} \label{5.8}
\E^\mu |\nabla u|_2^2 = \frac12 \BBBB_0, \quad   
\E^\mu |\Delta u|_2^2 \le \frac12 \BBBB_1\,,\quad
\BBBB_0^2/2\BBBB_1 \le \E^\mu |u|^2 \le \frac12 \BBBB_0\,,
\end{equation}
so $\mu$ is supported by the space~$H^2$. More delicate analysis of the convergence~\eqref{5.7} shows  that 
\begin{equation} \label{5.9}
\mu(K) =1,\quad \text{where}\quad K=\{u\in H^1: \rot u\in L_\infty\}\,,
\end{equation}
see \cite{GSV-2015}. Moreover, the measure~$\mu$ is invariant for the deterministic equation~\eqref{5.0}$|_{\nu=0}$, i.e. for the free 2d Euler equation
\begin{equation} \label{Euler}
\p_t u +(u\cdot\nabla) u+ \nabla p=0\,,\quad \diver u=0\,.
\end{equation}
See \cite{kuksin-jsp2004, KS-book}.\footnote{The measure~$\mu$ is supported by the space~$K$, on which the flow of the Euler equation is well defined in the sense of Yudovich and is continuous in the weak$^*$ topology of that set, see in~\cite{GSV-2015}. Before~\eqref{5.9} was obtained, an additional construction, suggested in \cite{kuksin-jsp2004},  was used  to explain in which sense the limiting measure~$\mu$ is invariant for~\eqref{Euler}.}  The limit~\eqref{5.7} is  the {\it inviscid limit} for the (properly scaled) stochastic 2d NSE. In view of what has been said at the beginning of the previous subsection, the inviscid limit measures $\mu$ describe the statistic of  space-periodic stationary 2d turbulence. We summarise the results concerning  this limit  in a theorem:

\begin{theorem}\label{t5.1}
{\bf(1)}\ 
Any inviscid limit   measure $\mu$  satisfies \eqref{5.8}, \eqref{5.9} and is invariant for Eq.~\eqref{Euler}.

\smallskip
\noindent
{\bf(2)}\ 
If $b_s\equiv b_{-s}$ and $|b_s|$ decay sufficiently fast as $|s|\to\infty$,  then the measure  $\mu$ is space-homogeneous and satisfies \eqref{5.6}, where $\Emn$ should be replaced by $\E^\mu$.  
\end{theorem}

The last assertion follows from \eqref{5.6}, convergence \eqref{5.7} and Fatou's lemma. Convergence~\eqref{5.7} does not allow to pass to the limit in  \eqref{5.4} and \eqref{5.5}, and we do not know if the balance relations hold for the inviscid limit measures~$\mu$. Relations~\eqref{5.8} imply that the limiting measures $\mu$ are non-trivial in the sense that they do not equal the delta-measure in the origin. In fact, they are non-degenerate in a much stronger sense. To state the corresponding result, we denote
$$
E(u) = \frac12 |u|_2^2,\quad
 E_1(u) = \frac12 |\rot u|_2^2= \frac12|\nabla u|_2^2,
$$
and call a real analytic function $f(r)$ {\it admissible} if $f''(r)$ has at most a polynomial growth as $|r|\to\infty$ and is bounded from below (e.g., $f$ is a polynomial of the form $f=r^{2m}+\cdots$, or any trigonometric polynomial).

\begin{theorem}[\cite{kuksin-cmp2008, KS-book}]\label{t5.2} 
{\bf(1)}\ 
The inviscid limits $\mu$ are such that the push-forward measures $E_*\mu$ and $(E_1)_* \mu$ are absolutely continuous with respect to the Lebesgue measure on~$\R$.

\smallskip
\noindent
{\bf(2)}\ 
If $\BBBB_2<\infty$ and $b_s\equiv b_{-s}$, then for any $d\in\N$ and any admissible functions $f_1,\dots, f_d$ such that their derivatives $f'_1,\dots, f'_d$ are linearly independent modulo constants,\,\footnote{That is, if  $c_1f'_1+\cdots+ c_df'_d\equiv \const$, then all $c_k$ vanish.} the push-forward of~$\mu$ under the mapping 
$$
u(\cdot ) \mapsto \Big(\int_{\T^2} f_k(\rot u(x))\,\dd x,\ 1\le k\le d\Big)\in\R^d\,,
$$
is absolutely continuous with respect to the Lebesgue measure. 
\end{theorem}

Due to this result, the Hausdorff dimension of~$\supp\mu$ is infinite. Indeed,  if this is not the case, then choosing~$d$ bigger than the Hausdorff dimension, we  see that the push-forward measure in item (2) of the theorem is supported by a set of dimension $<d$, which contradicts the assertion.

\section{3d Navier--Stokes system in thin domains}
\label{s6}
In this section we consider a thin layer around the torus~$\T^2$ and the 3d Navier--Stokes system, perturbed by a random kick-force with a sufficiently small vertical component. We show that when the width of the layer goes to zero, statistical characteristics of the 3d flow converge to those of the 2d flow \eqref{0.1}, \eqref{1.2}, \eqref{1.6}, where the kicks~$\eta^k$ are the horizontal components of the 3d kicks. Moreover, this convergence {\it is uniform in time}. Since Earth's atmosphere is a thin spherical layer, this result gives a good support to the belief that suitably chosen 2d stochastic meteorological models can be successfully used to model the climate. Usually these 2d models are related to Eq.~\eqref{0.1}; see the works~\cite{varner2013,klevtsova-2017} and the references therein.

Let $Q_\e = \T^2\times (0,\e) = \{x=(x_1, x_2, x_3)\}$. Consider the 3d Navier--Stokes system in $Q_\e$ 
under the free boundary conditions:
\begin{equation} \label{6.1}
\p_tu+\langle u,\nabla \rangle u-\nu\Delta u+\nabla p=\eta(t,x), \quad\diver u=0,\;\;
 x\in Q_\e\,,
\end{equation}
\begin{equation} \label{6.2}
u_3=\p_3 u_{1,2} =0\quad \text{for} \quad x_3=0\;\text{and}\; x_3=\e\,,
\end{equation}
\begin{equation} \label{6.3}
u(0,x) = u_0(x)\,,
\end{equation}
where $ u=(u_1,u_2,u_3)$. Denote by $H_\e$ (by $V_\e$) the $L_2$-space (the $H^1$-space) of divergence-free vector fields $(u_1,u_2,u_3)(x)$ on $Q_\e$ such that $u_1$ and $u_2$ have zero mean. Denote by $|\cdot |_\e$ the $L_2$-norm on $Q_\e$, i.e.~the norm in~$H_\e$ (note that $| \bf 1|_\e=\e)$, by $( \cdot,\cdot )_\e$  the corresponding~$L_2$ scalar product, and denote by $\|\cdot\|_{\e}$ the homogeneous norm in $V_\e$, $\|u\|_{\e} = |\nabla u |_\e$. The space~$H$ as in Section~\ref{s1.1} is naturally embedded in $H_\e$:
$$
i: H\ni (u_1,u_2) \mapsto (u_1(x_1, x_2), u_2(x_1, x_2),0) \in H_\e\,,
$$
and the  norm of this embedding equals $\sqrt\e$. Introduce in $H_\e$ two orthogonal projections:
$$
M_\e u =( \e^{-1} \int_0^\e u_1(x',y)\,\dd y, \e^{-1} \int_0^\e u_2(x',y)\,\dd y, 0),\quad
N_\e = \text{id}- M_\e\,,
$$
where $x'=(x_1,x_2)$. Then $M_\e H_\e = i H$. The 3d Stokes operator $L_\e = -\Delta|_{H_\e}$  preserves the spaces $M_\e H_\e$ and $N_\e H_\e$, and its eigenfunctions are of two kinds:
$$
e_s(x_1,x_2) \in M_\e H_\e,\; s\in\Z^2_\e\,,\quad\text{and}\quad e_j^\e(x_1, x_2, x_3)\in N_\e H_\e, \; j\in\N\,,
$$
where $|e_s|_\e = |e_j^\e|_\e=\sqrt\e$ for all $s$ and $j$, and
$$
L_\e e_s = |s|^2 e_s\; \;\forall s,\qquad L_\e e_j^\e = \Lambda_j^\e e_j^\e\;\; \forall\,j\,,
$$
so that 
$$
\|e_s\|_\e = ( Le_s, e_s)_\e^{1/2} = |s| \sqrt\e,\quad \| e^\e_j\|_\e = \sqrt{\Lambda_j^\e \,\e}\,.
$$
The  vectors $e_s \in i H$ will be  identified with the eigenvectors of the 2d Stokes operator (which is the opposite of the 2d Laplacian, restricted to the space~$H$). Each vector $e_j^\e$ has components 
$(e_j^\e)_l=
C_j^l (\sin/\cos)(s_j^l\cdot x')\,\cos(\frac\pi{\e} n_j^lx_3)$, $ l=1,2,3,
$
where  $s_j^l\in \Z^2_*$,  $n_j^l\in\N\cup\{0\}$ and at least one of the numbers $n_j^1,\dots,n_j^3$ is non-zero. Therefore the eigenvalue $\Lambda_j^\e$ has the form 
\begin{equation}  \label{6.}
\Lambda_j^\e = A_j + B_j\pi^2 \e^{-2},\quad A_j\in \N\cup\{0\}, \; B_j\in\N\,.
\end{equation}

Assume that the force $\eta(t,x)$ in \eqref{6.1} is a kick-process of the form \eqref{1.6} with the kicks 
$$
\eta^k_\e(x) = \sum_{s\in\Z^2_*} b_s \xi^k_s e_s(x) + \sum_{j=1}^\infty d_j^\e \zeta^k_je^\e_j(x)\,.
$$
Here the constants $\{b_s\}$ and $\{d^\e_j\}$ are such that 
\begin{equation} \label{6.4}
\BBBB_1:= \sum b_s^2|s|^2<\infty,\;\;\; b_s\ne0 \; \forall\, s, \quad
\DD^\e_1:= \sum(d^\e_j)^2 \Lambda_j^\e <\infty\,,
\end{equation}
and $\{\xi^k_s\}$, $\{\zeta^k_{j}\}$ are i.i.d. random variable with law $p(r)\dd r$, where $p\in C_0^1(-1,1)$ satisfies the conditions
$$
p(0)\ne0,\quad \int_\R rp(r)\,\dd r=0.
$$ 

We shall compare solutions of Eqs.~\eqref{6.1}--\eqref{6.3} with those of the kick-forced  2d Navier--Stokes system \eqref{0.1}, \eqref{1.2}, \eqref{1.6}, where the kicks $\eta^k$ are
$$
\eta^k = M_\e \eta^k_\e=\sum_s b_s \zeta^k_s e_s(x)\,.
$$
Under the assumptions \eqref{6.4}, this 2d equation is exponentially mixing in the space $H$ in the sense that
there exists a measure $\mu\in\PP(H)$ such that for every solution $u(t)$ of \eqref{0.1}, \eqref{1.2}, \eqref{1.6} its law,
evaluated at  integer points $t\in\N$, converges to $\mu$ exponentially fast in the dual-Lipschitz norm in $\PP(H)$; see \cite{KS-book}. The mixing property for Eqs.~\eqref{6.1}--\eqref{6.3}, claimed in the theorem 
below, is understood in a similar way. 

\begin{theorem}\label{t6.1}
Assume that 
\begin{equation} \label{6.5}
\DD_1^\e \le \e^{-1} \gamma^2(\e),\quad \text{where}\quad \gamma(\e)\to0\;\; \text{as}\;\; \e \to0\,.
\end{equation}
Then there exist $c_0, \e_0>0$ such that the following properties hold.
\begin{description}
\item[Exponential mixing.] 
If $0<\e\le\e_0$, then the set
$$
\OO_\e = \{ u: \|M_\e u\|_\e \le c_0\sqrt\e,\; \|N_\e u\|_\e \le  c_0 \gamma(\e)\} \subset V_\e
$$
is invariant for Eqs.~\eqref{6.1}--\eqref{6.3} and  the dynamics on $\OO_\e$ is exponentially mixing with invariant measure $\mu_\e\in\PP(\OO_\e)$. 

\item[Convergence.]
As $\e\to0$, the measure $(M_\e)_* \mu_\e$ converges to $ \mu$ weakly in $H$.

\item[Stability.]
Let $v_0\in H$ be such that $\|v_0\|_1< c_0$, let $u(t)$ be a solution of \eqref{0.1}, \eqref{1.2}, \eqref{1.6}, equal~$v_0$ at $t=0$, and let~$u_\e(t)$ be a solution of Eqs.~\eqref{6.1}--\eqref{6.3} with $u_0=iv_0 \in \OO_\e$. Then 
$$
\|\DD(M_\e u_\e(t)) - \DD(u(t))\|^*_{L(H)} \to 0 \quad \text{as}\quad \e\to 0,
$$
uniformly in $t\ge0$. 
\end{description}
\end{theorem}

For a random vector field $u$ on $Q_\e$ its averaged normalised kinetic energy is $\EE_\e (u) = \tfrac12 \E\,|u|^2_\e /\Vol (Q_\e)=\frac1{2\e}\E\,|u|^2_\e$. By \eqref{6.5} and \eqref{6.}, for a kick $\eta^k_\e$, we have
$$
\EE_\e(M_\e\eta^k_\e) = \frac12\kappa^2 \sum_s b_s^2 \sim1, 
\quad 
\EE_\e(N_\e\eta^k_\e) = \frac12\kappa^2 \sum_j (d_j^\e)^2 \lesssim \e\gamma^2(\e)
$$
where  $\kappa^2 = \int r^2p(r)\,\dd r$. For the averaged normalised dissipation of energy $ \DD_\e(u) = \frac1{2\e}\E\,\|u\|_\e^2$, we have 
$$
\DD_\e(M_\e\eta^k_\e) = \frac12 \kappa^2 \BBBB_1\sim1, \quad \DD_\e(N_\e\eta^k_\e) = \tfrac12 \kappa^2 \DD_1^\e  \le\frac12\kappa^2\e^{-1} \gamma^2(\e).
$$ 
Therefore the vertical component of the random force in~\eqref{6.1} should be small in terms of the energy, but not in terms of the dissipation of energy.

\section{Open problems}
\label{s7}

\begin{OP}[Mixing of pipe flow]
Let us consider the Navier--Stokes system~\eqref{0.1} in the strip
$$
D=\{(x_1,x_2)\in\R^ 2: x_1\in\R, |x_2|<1\}.
$$
The initial-boundary value problem for~\eqref{0.1} is well posed in~$L^\infty$ spaces on~$D$, with no decay conditions at infinity (see~\cite{AM-2005,AZ-2014}). Moreover, the problem is dissipative and possesses a global attractor (in the deterministic setting). It follows that, at least in the case of a bounded stochastic forcing, the random dynamics has a stationary distribution. A challenging problem is to prove its {\it uniqueness\/} and {\it mixing\/}. 
\end{OP}

\begin{OP}[Mixing in the whole space]
The 2d Navier--Stokes system considered on the whole space~$\R^2$ is well posed in $L^\infty$ spaces (see~\cite{GMS-2001,lr2002}). However, it is not known if the dynamics is dissipative, and it seems to be a hopeless task to prove any kind of regular behaviour of solutions under stochastic perturbations. On the other hand, the Navier--Stokes system with the 
 Ekman damping 
\begin{equation} \label{7.1}
\p_tu+\gamma u+\langle u,\nabla \rangle u-\nu\Delta u+\nabla p=f(t,x), \quad\diver u=0,
\end{equation}
where $\gamma>0$ is a number, is dissipative (see~\cite{zelik-2013}). It is a natural question to investigate the {\it existence of stationary measure\/} and its {\it mixing properties\/} under various types of random perturbation.
\end{OP}

\begin{OP}[Vanishing noise limit]\label{op3}
The vanishing noise limit of stationary measures described in Theorem~\ref{t4.2} concerns a rather particular situation: the unperturbed dynamics should be globally asymptotically stable. This is a very restrictive hypothesis, and removing it is an important question. In the finite-dimensional case this problem is rather well understood, and one can establish the so-called {\it Freidlin--Wentzell asymptotics\/} for stationary measures; see Section~6.4 in~\cite{FW2012}. As for stochastic PDE's, similar results can be proved, provided that the global attractor for the unperturbed dynamics has a regular structure. The latter means that the attractor consists of finitely many steady states and the heteroclinic orbits joining them. Such a result is proved in~\cite{martirosyan-2017} for the case of a damped nonlinear wave equation. The attractor of the Navier--Stokes system is not likely to possess that property, and the validity of the {\it Freidlin--Wentzell type asymptotics\/} for stationary distributions remains an open problem. 
\end{OP}

\begin{OP}[Inviscid limit]\label{op4}
{\bf(a)}\ 
Does the limiting measure $\mu$ in \eqref{5.7} depend on the sequence $\{\mu_{\nu_j}\}$?  (We believe that it does). 

\smallskip
{\bf(b)}\ 
Consider the energy spectrum of a mesure $\mu$ as above:
$$
E_k(\mu) = Z^{-1} \sum_{\{s\in\Z^2_*: M^{-1} k\le |s| \le Mk\}} \E^\mu |u_s|^2,
$$
where $Z$ is the number of  terms in the sum and $M>1$ is a suitable constant. Do there exist positive constants $a,b,C$ such that 
\begin{equation} \label{7.4}
C^{-1} k^{-a} \le E_k(\mu) \le Ck^{-b}
\end{equation}
for all $k\ge1$? Do the exponents $a$ and $b$ depend on the inviscid limit $\mu$?

\smallskip
{\bf(c)}\ 
Do the stationary measures $\mu^\nu$ satisfy~\eqref{7.4} for $C_1 \nu^{-\alpha} \le k\le C_2\nu^{-\beta}$, for suitable $0\le \alpha<\beta$ and $C_1, C_2>0$?  See \cite{boritchev-2013} for an affirmative answer to this question with $a=b=2$ and $\alpha=0, \beta=1$, when $u$ is a solution of the 1d stochastic Burgers equation. 
\end{OP}

\begin{OP}[thin 3d domains]\label{op5}
Improve the result of Theorem \ref{t6.1} by replacing condition~\eqref{6.5} with a weaker restrain, thus allowing in \eqref{6.1} for random forces with bigger vertical components (this seems to be possible to achieve by making better use of the stochastic nature of the force $\eta$). Obtain similar results for 3d stochastic models of  Earth's atmosphere and their suitable 2d approximations. 
\end{OP}

%\newpage
\subsection*{Frequently used notations}

\begin{tabular}{p{2.7cm}p{8.9cm}}
$C$, $C_i$ & positive numbers that may depend on the parameters mentioned in brackets\\

$J\subset\R$ & closed interval\\

$C(J,E)$ & space of continuous functions on~$J$ with range in a Banach space~$E$\\

$L^p(J,E)$ & space of Borel-measurable functions $f:J\to E$ such that $\int_J\|f(t)\|_E^p\,\dd t<\infty$, with obvious modification for $p=\infty$\\ [2pt]

$X$ & complete separable metric space with a distance~$d_X$\\

$\BB(X)$ & Borel $\sigma$-algebra of~$X$\\

$C_b(X)$ & space of bounded continuous functions $g:X\to\R$ with the norm
$
\|g\|_\infty=\sup_{u\in X}|g(u)|
$\\

$d_X(u,B)$ & minimal distance from a point $u\in X$ to a subset $B\subset X$\\

$\PP(X)$ & space of Borel probability measures on~$X$ endowed with the topology of weak convergence; the latter can be metrised by the Lipschitz-dual metric $\|\cdot\|^*_{\text{Lip}(X)}$, see \eqref{LD}\\

$\DD(\xi)$ & the law of a random variable~$\xi$\\

$F_*\mu$ & the image of a measure~$\mu$ under a measurable mapping~$F$\\

$L^p(D)$, $H^s(D)$ & Lebesgue and Sobolev spaces on a domain~$D\subset\R^d$ with standard norms~$|\cdot|_p$ and~$\|\cdot\|_s$, respectively; sometimes  we write  $L^p(D,\R^m)$ and $H^s(D,\R^m)$ to denote the corresponding spaces of $\R^m$-valued functions, and~$(\cdot,\cdot)$ stands for $L_2$-scalar product\\ 

$H^s$  & space of divergence-free vector functions with zero mean that belong to $H^s(\T^2,\R^2)$, $s\in\N\cup\{0\}$.
\\

$H$, $V$ &  abbreviations for the spaces $H^0$ and $H^1$\\

$L^2_\sigma$ & the set of divergence--free functions in $L^2(\T^2,\R^2)$\\

$\R_+$ , $\Z_+$ & sets of real numbers and non-negative integers, respectively\\

$\T^2=\R^2/2\pi\Z^2$ & two-dimensional torus with sides~$2\pi$\\

$(\Omega,\FF,\IP)$ & complete probability space\\
\end{tabular}

%\newpage
\addcontentsline{toc}{section}{Bibliography}
\def\cprime{$'$} \def\cprime{$'$}
  \def\polhk#1{\setbox0=\hbox{#1}{\ooalign{\hidewidth
  \lower1.5ex\hbox{`}\hidewidth\crcr\unhbox0}}}
  \def\polhk#1{\setbox0=\hbox{#1}{\ooalign{\hidewidth
  \lower1.5ex\hbox{`}\hidewidth\crcr\unhbox0}}}
  \def\polhk#1{\setbox0=\hbox{#1}{\ooalign{\hidewidth
  \lower1.5ex\hbox{`}\hidewidth\crcr\unhbox0}}} \def\cprime{$'$}
  \def\polhk#1{\setbox0=\hbox{#1}{\ooalign{\hidewidth
  \lower1.5ex\hbox{`}\hidewidth\crcr\unhbox0}}} \def\cprime{$'$}
  \def\cprime{$'$} \def\cprime{$'$} \def\cprime{$'$}
\providecommand{\bysame}{\leavevmode\hbox to3em{\hrulefill}\thinspace}
\providecommand{\MR}{\relax\ifhmode\unskip\space\fi MR }
\providecommand{\MRhref}[2]{%
  \href{http://www.ams.org/mathscinet-getitem?mr=#1}{#2}
}
\providecommand{\href}[2]{#2}


\begin{thebibliography}{JNPS17}

\bibitem[AM05]{AM-2005}
A.~L. Afendikov and A.~Mielke, \emph{Dynamical properties of spatially
  non-decaying 2{D} {N}avier-{S}tokes flows with {K}olmogorov forcing in an
  infinite strip}, J. Math. Fluid Mech. \textbf{7} (2005), no.~suppl. 1,
  S51--S67.

\bibitem[AZ14]{AZ-2014}
P.~Anthony and S.~Zelik, \emph{Infinite-energy solutions for the
  {N}avier-{S}tokes equations in a strip revisited}, Commun. Pure Appl. Anal.
  \textbf{13} (2014), no.~4, 1361--1393.

\bibitem[Bat82]{batchelor1982}
G.~K. Batchelor, \emph{The {T}heory of {H}omogeneous {T}urbulence}, Cambridge
  University Press, Cambridge, 1982.

\bibitem[BC17]{BC-2017}
Z.~Brze{\'z}niak and S.~Cerrai, \emph{{Large deviation principle for the
  invariant measures of the 2D stochastic Navier--Stokes equations on a
  torus}}, J. Funct. Anal. \textbf{273} (2017), no.~6, 1891 -- 1930.

\bibitem[BKL02]{BKL-2002}
J.~Bricmont, A.~Kupiainen, and R.~Lefevere, \emph{Exponential mixing of the
  2{D} stochastic {N}avier--{S}tokes dynamics}, Comm. Math. Phys. \textbf{230}
  (2002), no.~1, 87--132.

\bibitem[Bor13]{boritchev-2013}
A.~Boritchev, \emph{{Sharp estimates for turbulence in white-forced generalised
  Burgers equation}}, Geom. Funct. Anal. \textbf{23} (2013), no.~6, 1730--1771.

\bibitem[BV12]{BV-2012}
F.~Bouchet and A.~Venaille, \emph{Statistical mechanics of two-dimensional and
  geophysical flows}, Phys. Rep. \textbf{515} (2012), no.~5, 227--295.

\bibitem[CM10]{CM-2010}
I.~Chueshov and A.~Millet, \emph{{Stochastic 2D hydrodynamical type systems:
  Wellposedness and large deviations}}, Appl. Math. Optim. \textbf{61} (2010),
  no.~3, 379--420.

\bibitem[Cra91]{crauel-1991}
H.~Crauel, \emph{Markov measures for random dynamical systems}, Stochastics
  Stochastics Rep. \textbf{37} (1991), no.~3, 153--173.

\bibitem[DV75]{DV-1975}
M.~D. Donsker and S.~R.~S. Varadhan, \emph{{Asymptotic evaluation of certain
  Markov process expectations for large time, I-II}}, Comm. Pure Appl. Math.
  \textbf{28} (1975), 1--47, 279--301.

\bibitem[EMS01]{EMS-2001}
W.~E, J.~C. Mattingly, and Ya. Sinai, \emph{Gibbsian dynamics and ergodicity
  for the stochastically forced {N}avier--{S}tokes equation}, Comm. Math. Phys.
  \textbf{224} (2001), no.~1, 83--106.

\bibitem[Fla94]{flandoli-1994}
F.~Flandoli, \emph{Dissipativity and invariant measures for stochastic
  {N}avier--{S}tokes equations}, NoDEA Nonlinear Differential Equations Appl.
  \textbf{1} (1994), no.~4, 403--423.

\bibitem[FM95]{FM-1995}
F.~Flandoli and B.~Maslowski, \emph{Ergodicity of the 2{D} {N}avier--{S}tokes
  equation under random perturbations}, Comm. Math. Phys. \textbf{172} (1995),
  no.~1, 119--141.

\bibitem[FP67]{FP-1967}
C.~Foia{\c{s}} and G.~Prodi, \emph{Sur le comportement global des solutions
  non-stationnaires des \'equations de {N}avier--{S}tokes en dimension {$2$}},
  Rend. Sem. Mat. Univ. Padova \textbf{39} (1967), 1--34.

\bibitem[Fri95]{frisch1995}
U.~Frisch, \emph{Turbulence. {T}he {L}egacy of {A}. {N}. {K}olmogorov},
  Cambridge University Press, Cambridge, 1995.

\bibitem[FW12]{FW2012}
M.~I. {Freidlin} and A.~D. {Wentzell}, \emph{{Random Perturbations of Dynamical
  Systems}}, Springer, Heidelberg, 2012.

\bibitem[Gal02]{gallavotti2002}
G.~Gallavotti, \emph{Foundations of {F}luid {D}ynamics}, Springer-Verlag,
  Berlin, 2002.

\bibitem[GMS01]{GMS-2001}
Y.~Giga, S.~Matsui, and O.~Sawada, \emph{Global existence of
  two-dimen\-sio\-nal {N}avier--{S}tokes flow with nondecaying initial
  velocity}, J. Math. Fluid Mech. \textbf{3} (2001), no.~3, 302--315.

\bibitem[GSV15]{GSV-2015}
N.~{Glatt-Holtz}, V.~Sverak, and V.~Vicol, \emph{On inviscid limits for the
  stochastic {Navier}--{S}tokes equations and related models}, Arch. Rat. Mech.
  Anal. \textbf{217} (2015), no.~1, 619--649.

\bibitem[HM06]{HM-2006}
M.~Hairer and J.~C. Mattingly, \emph{Ergodicity of the 2{D} {N}avier--{S}tokes
  equations with degenerate stochastic forcing}, Ann. of Math. (2) \textbf{164}
  (2006), no.~3, 993--1032.

\bibitem[HM11]{HM-2011}
\bysame, \emph{A theory of hypoellipticity and unique ergodicity for semilinear
  stochastic {PDE}s}, Electron. J. Probab. \textbf{16} (2011), no. 23,
  658--738.

\bibitem[JNPS15]{JNPS-cpam2015}
V.~{Jak\v si\'c}, V.~Nersesyan, C.-A. Pillet, and A.~Shirikyan, \emph{Large
  deviations from a stationary measure for a class of dissipative {PDE}s with
  random kicks}, Comm. Pure Appl. Math. \textbf{68} (2015), no.~12, 2108--2143.

\bibitem[JNPS17]{JNPS-2014}
\bysame, \emph{{Large deviations and mixing for dissipative PDE's with
  unbounded random kicks}}, Nonlinearity (2017), to appear.

\bibitem[Kle17]{klevtsova-2017}
Yu.~Yu. Klevtsova, \emph{On the rate of convergence of distributions of
  solutions to the stationary measure as {$t\to+\infty$} for the stochastic
  system of the {L}orenz model describing a baroclinic atmosphere}, Mat. Sb.
  \textbf{208} (2017), no.~7, 19--67.

\bibitem[KNS18]{KNS-2017}
S.~Kuksin, V.~Nersesyan, and A.~Shirikyan, \emph{{Exponential mixing for a
  class of dissipative PDEs with bounded degenerate noise}}, in preparation
  (2018).

\bibitem[KP05]{KP-jsp2005}
S.~Kuksin and O.~Penrose, \emph{A family of balance relations for the
  two-dimensional {N}avier--{S}tokes equations with random forcing}, J.
  Statist. Phys. \textbf{118} (2005), no.~3-4, 437--449.

\bibitem[KS91]{KS1991}
I.~Karatzas and S.~E. Shreve, \emph{Brownian {M}otion and {S}tochastic
  {C}alculus}, Springer-Verlag, New York, 1991.

\bibitem[KS00]{KS-cmp2000}
S.~Kuksin and A.~Shirikyan, \emph{Stochastic dissipative {PDE}s and {G}ibbs
  measures}, Comm. Math. Phys. \textbf{213} (2000), no.~2, 291--330.

\bibitem[KS01]{KS-cmp2001}
\bysame, \emph{A coupling approach to randomly forced nonlinear {PDE}'s. {I}},
  Comm. Math. Phys. \textbf{221} (2001), no.~2, 351--366.

\bibitem[KS02]{KS-jmpa2002}
\bysame, \emph{Coupling approach to white-forced nonlinear {PDE}s}, J. Math.
  Pures Appl. (9) \textbf{81} (2002), no.~6, 567--602.

\bibitem[KS03]{KS-prse2003}
\bysame, \emph{Some limiting properties of randomly forced two-dimensional
  {N}avier--{S}tokes equations}, Proc. Roy. Soc. Edinburgh Sect. A \textbf{133}
  (2003), no.~4, 875--891.

\bibitem[KS04]{KS-faa2004}
\bysame, \emph{On random attractors for systems of mixing type}, Funktsional.
  Anal. i Prilozhen. \textbf{38} (2004), no.~1, 34--46, 95.

\bibitem[KS12]{KS-book}
\bysame, \emph{Mathematics of {T}wo-{D}imensional {T}urbulence}, Cambridge
  University Press, Cambridge, 2012.

\bibitem[Kuk02a]{kuksin-rmp2002}
S.~Kuksin, \emph{Ergodic theorems for 2{D} statistical hydrodynamics}, Rev.
  Math. Phys. \textbf{14} (2002), no.~6, 585--600.

\bibitem[Kuk02b]{kuksin-ams2002}
\bysame, \emph{On exponential convergence to a stationary measure for nonlinear
  {PDE}s perturbed by random kick-forces, and the turbulence limit}, Partial
  differential equations, Amer. Math. Soc. Transl. Ser. 2, vol. 206, Amer.
  Math. Soc., Providence, RI, 2002, pp.~161--176.

\bibitem[Kuk04]{kuksin-jsp2004}
\bysame, \emph{The {E}ulerian limit for 2{D} statistical hydrodynamics}, J.
  Statist. Phys. \textbf{115} (2004), no.~1-2, 469--492.

\bibitem[Kuk08]{kuksin-cmp2008}
\bysame, \emph{On distribution of energy and vorticity for solutions of 2{D}
  {N}avier--{S}tokes equation with small viscosity}, Comm. Math. Phys.
  \textbf{284} (2008), no.~2, 407--424.

\bibitem[Lam96]{lamperti1996}
J.~Lamperti, \emph{Probability}, John Wiley \& Sons, New York, 1996.

\bibitem[{Le }87]{lejan-1987}
Y.~{Le Jan}, \emph{\'{E}quilibre statistique pour les produits de
  diff\'eo\-mor\-phismes al\'eatoires ind\'ependants}, Ann. Inst. H. Poincar\'e
  Probab. Statist. \textbf{23} (1987), no.~1, 111--120.

\bibitem[Led86]{ledrappier-1986}
F.~Ledrappier, \emph{Positivity of the exponent for stationary sequences of
  matrices}, Lyapunov {E}xponents (Bremen, 1984), Springer, Berlin, 1986,
  pp.~56--73.

\bibitem[LR02]{lr2002}
P.-G. Lemari{\'e}-Rieusset, \emph{Recent {D}evelopments in the
  {N}avier--{S}tokes {P}roblem}, Chapman \& Hall/CRC, Boca Raton, FL, 2002.

\bibitem[Mar17a]{martirosyan-2016}
D.~Martirosyan, \emph{{Large deviations for invariant measures of white-forced
  2D Navier--Stokes equation}}, J. Evol. Equ. (2017), accepted for publication.

\bibitem[Mar17b]{martirosyan-2017}
\bysame, \emph{Large deviations for stationary measures of stochastic nonlinear
  wave equations with smooth white noise}, Comm. Pure Appl. Math. \textbf{70}
  (2017), no.~9, 1754--1797.

\bibitem[Mat02]{mattingly-2002}
J.~C. Mattingly, \emph{Exponential convergence for the stochastically forced
  {N}avier--{S}tokes equations and other partially dissipative dynamics}, Comm.
  Math. Phys. \textbf{230} (2002), no.~3, 421--462.

\bibitem[MN17]{MN-2015}
D.~Martirosyan and V.~Nersesyan, \emph{Local large deviations principle for
  occupation measures of the damped nonlinear wave equation perturbed by a
  white noise}, Ann. Inst. H. Poincar\'e Probab. Statist. (2017), to appear.

\bibitem[Ner17]{nersesyan-2017}
V.~Nersesyan, \emph{{Large deviations for the Navier--Stokes equations driven
  by a white-in-time noise}}, in preparation (2018).

\bibitem[Oda08]{odasso-2008}
C.~Odasso, \emph{Exponential mixing for stochastic {PDE}s: the non-additive
  case}, Probab. Theory Related Fields \textbf{140} (2008), no.~1-2, 41--82.

\bibitem[Shi05]{shirikyan-rjmp2005}
A.~Shirikyan, \emph{Ergodicity for a class of {M}arkov processes and
  applications to randomly forced {PDE}'s. {I}}, Russ. J. Math. Phys.
  \textbf{12} (2005), no.~1, 81--96.

\bibitem[Shi06]{shirikyan-ptrf2006}
\bysame, \emph{Law of large numbers and central limit theorem for randomly
  forced {PDE}'s}, Probab. Theory Related Fields \textbf{134} (2006), no.~2,
  215--247.

\bibitem[Shi15]{shirikyan-asens2015}
\bysame, \emph{Control and mixing for 2{D} {N}avier-{S}tokes equations with
  space-time localised noise}, Ann. Sci. \'Ec. Norm. Sup\'er. (4) \textbf{48}
  (2015), no.~2, 253--280.

\bibitem[Shi18]{shirikyan-2017}
\bysame, \emph{{Controllability implies mixing II. Convergence in the
  dual-Lipschitz metric}}, in preparation (2018).

\bibitem[Tem79]{temam1979}
R.~Temam, \emph{Navier--{S}tokes {E}quations}, North-Holland, Amsterdam, 1979.

\bibitem[Var13]{varner2013}
G.~A. Varner, \emph{Stochastically {P}erturbed {N}avier--{S}tokes {S}ystem on
  the {R}otating {S}phere}, ProQuest LLC, Ann Arbor, MI, 2013.

\bibitem[VF88]{VF1988}
M.~I. Vishik and A.~V. Fursikov, \emph{{M}athematical {P}roblems in
  {S}tatistical {H}ydromechanics}, Kluwer, Dordrecht, 1988.

\bibitem[Zel13]{zelik-2013}
S.~Zelik, \emph{Infinite energy solutions for damped {N}avier-{S}tokes
  equations in~{$\Bbb{R}^2$}}, J. Math. Fluid Mech. \textbf{15} (2013), no.~4,
  717--745.
\end{thebibliography}
\end{document}